\documentclass[12pt,a4paper]{article}

\usepackage{graphicx}

\newcommand{\dd}{\ensuremath\mathrm{d}}
\newcommand{\EF}{\ensuremath E_\mathrm{F}}
\newcommand{\eL}{\ensuremath \epsilon_{_\mathrm{L}}}
\newcommand{\eS}{\ensuremath \epsilon_{_\mathrm{S}}}
\newcommand{\eR}{\ensuremath \epsilon_{_\mathrm{R}}}
\newcommand{\KL}{\ensuremath \mathcal{K}_\mathrm{L}}
\newcommand{\KS}{\ensuremath \mathcal{K}_\mathrm{S}}
\newcommand{\KR}{\ensuremath \mathcal{K}_\mathrm{R}}
\newcommand{\tB}{\ensuremath t_\mathrm{B}}
\newcommand{\Nb}{\ensuremath N_\mathrm{b}}
\newcommand{\tBS}{\ensuremath t_\mathrm{BS}}
\newcommand{\mmax}{\ensuremath m_\mathrm{max}}
\newcommand{\lmax}{\ensuremath l_\mathrm{max}}
\newcommand{\tsw}{\ensuremath t_\mathrm{sw}}

\usepackage{amssymb,amsmath}
\usepackage{ifthen}

\newboolean{twocol}
\setboolean{twocol}{false}

\begin{document}
\title{Stroboscopic wave packet description of 
time-dependent currents through ring-shaped nanostructures}%
\author{%
	Martin Kon\^{o}pka$^{1}$
	\and
	Peter Bokes$^{1,2}$
	\\ \\
\and	$^1$Department of Physics\\
	Institute of Nuclear and Physical Engineering\\
	Faculty of Electrical Engineering and Information Technology\\
	Slovak University of Technology in Bratislava\\
	Ilkovi\v{c}ova 3, 812~19 Bratislava, Slovakia
\and	$^2$European Theoretical Spectroscopical Facility\\
	(ETSF, \texttt{www.etsf.eu})%
}
%
\date{}
\maketitle
\begin{abstract}
We present an implementation of a new method for explicit simulations
of time-dependent electric currents through nanojunctions.
The method is based on unitary propagation of stroboscopic wave packet
states and is designed to treat open systems with fluctuating number of
electrons while preserving full quantum coherence throughout the whole infinite system.
We demonstrate the performance of the method on a model system
consisting of a ring-shaped nanojunction with two semi-infinite
tight-binding leads.
Time-dependent electron current responses to abrupt bias turn-on or
gate potential switching are computed for several ring configurations
and ring-leads coupling parameters.
The found current-carrying stationary states agree well
with the predictions of the Landauer formula.
As examples of genuinely time-dependent process we explore
the presence of circulating currents in the rings in transient regimes
and the effect of a time-dependent gate potential.
\end{abstract}
%
%
\section{Introduction}
%
%
Transition from the state with zero electric current to a state with nonzero current 
in nanoelectronic devices is a technologically important example of non-equilibrium 
quantum dynamics of an open many-electron system.
It exhibits several significantly different 
time scales related to relaxation of electrons with important consequences
for the prospects of their applications in nanoelectronics.
The common problem in theoretical quantum transport description 
in both stationary and time-dependent situations is proper
inclusion of the semi-infinite leads.
The number of degrees of freedom in the nanojunction, which are explicitly 
considered, has to be sufficiently small in order to be numerically tractable.
Hence the boundary conditions at the junction must be correctly described
and included.
An approach widely used for stationary systems are the non-equilibrium Green's functions (NEGF) with lead 
self-energies~\cite{NEGF,Ness11} which include the physics induced by the leads.

The description of boundary conditions of the finite nanojunction becomes more complicated 
in time-dependent problems. The non-stationary quantum transport has been addressed 
for non-interacting electrons~\cite{Cini80}, within the time-dependent density-functional theory 
(TDDFT)~\cite{Stefanucci04,TDDFT} or even the many-body perturbation theory~\cite{MBPT}.
It has been shown that self-energies -- now time-dependent -- could in principle
be used also in this case as long as the leads are treated at the TDDFT level.
In practice, it is a quite complicated approach especially for realistically
structured systems so that simplifications of the formalism are necessary~\cite{BWang10,YWang11}.
Chen \textit{et al.} proved the so-called holographic electron density
theorem and developed a new method for time-dependent
open electronic systems~\cite{Chen07}.
Further effort in this direction led to computationally more efficient 
density-functional tight-binding method~\cite{YWang11}.
Very recent works of Chen \textit{et al.} on time-dependent quantum transport are based
on Liouville-von-Neumann equation for single-electron density matrix~\cite{Chen12}.
Another recent work by a different group~\cite{Torfason12} uses generalised master equation approach
to mesoscopic time-dependent transport.
Non-equilibrium thermodynamical theory of interacting tunnelling transport 
has been presented by Hyldgaard~\cite{Hyldgaard12}.

Another class of methods to tackle time-dependent transport with open boundary conditions
has been in development~\cite{Horsfield04,McEniry07,McEniry10}.
The scope of these methods is wider than just elastic transport.
The methods are know as correlated electron-ion dynamics (CEID).
The dynamics is based on Ehrenfest molecular dynamics and its extensions.
The electronic degrees of freedom in CEID are divided between the central
system and its environment which is facilitated by the formalism
of one-particle density matrices.
A damping term has been introduced used in the open-boundary equations of motion~\cite{Horsfield04};
this term keeps the environment close to a reference state.
In further development of the method~\cite{McEniry07}, open boundaries have
been introduced in a new way which represents an explicit realisation of an external 
battery.
We remark that this method does not conserve
coherence between injection and subsequent scattering of the electrons.

In the present work we address the open-boundaries time-dependent quantum transport problem
using the stroboscopic wave packet (SWP) basis method, principles of which has been
provided in the works~\cite{bokes_PRL,bokes_PCCP}.
We study time-dependent currents through
model-system nanojunctions formed by small rings inserted to electric circuit
formed by mono-atomic leads.
We also present new developments of the stroboscopic wave packet approach (SWPA)
which are necessary in order to describe the systems with atomistic structure.
While in the orginal works~\cite{bokes_PRL,bokes_PCCP} structureless electrodes
and very simple tunnelling barriers have been used to demonstrate the method,
here we work with atomistic models of the electrodes and more complex 
nanojunctions.
The formulation presented here solves
the time-dependent Schr\"odinger equation (SchE) for independent electrons within 
the tight-binding (TB) approximation.
SchE is solved numerically with state vectors expanded in the
stroboscopic wave packet basis representation (SWB)~\cite{bokes_PRL,bokes_PCCP}.
The SWPA has been designed to be specifically
suited for time-dependent transport through nanojunctions.
Its main advantage is to make possible explicit integration of
equations of motions of nanojunction degrees of freedom
with full quantum coherence preserved throughout the whole infinite system.
This can be viewed also as an open-system treatment with correct inclusion
of boundary conditions.
In the present paper we focus our study on simple model
cases of rings and monoatomic leads with possible extension to realistic
systems.
Our main interest are transient currents developed in response to
abruptly applied bias and time-dependent gate potential.
We also refer to stationary results obtained by other authors
using analytical methods.

Transport properties of atomic-scale sized rings have been studied by several authors
in most cases on the H\"{u}ckel/tight-binding level of description
and in the stationary regime.
Effect of the asymmetric position of lead on transmittance has been studied
in Ref.~\cite{Yi03} by means of Green's functions.
Authors of work~\cite{Goyer07} have developed a source-sink
potential method for convenient analytical treatment of molecular
electronic devices including conjugated systems.
This approach has been used in Ref.~\cite{Pickup08} to obtain
the form  of the transmission with explicitly given dependence on the
molecular skeleton and its connection to leads.
Authors of Ref.~\cite{Stefanucci09} generalised the so-called
waveguide approach~\cite{Xia92} (WGA) to systems described by the
TB approximation.
Resulting methodology -- the tight-binding waveguide approach
(TBWGA) -- can be interpreted as a route to an exact solution of the stationary SchE
for given class of TB models.
An explicit formula for transmittance 
has been derived~\cite{Stefanucci09} for rings
composed of identical atoms and identical nearest-neighbour couplings
and coupled to two leads with equal chemical potentials in both of them.
The authors discussed conditions for occurrence of circulating currents
in such quantum rings.
More recently Sparks \textit{et al.} formulated the stationary
problem more generally,
considering a multibranch device in a TB approximation~\cite{SGML11}.
They provided an explicit formula for system transmittance
valid for a large class of systems including those with
different chemical potentials in different leads.
The formula has been obtained by an exact solution of 
the stationary SchE for the whole system
and has also been related to Green's function analysis.
Explicit results in Ref.~\cite{SGML11} facilitate 
the study of quantum interference
effects in stationary regime for wide class of systems.
We refer to these results especially when the long-time behaviour 
of our time-dependent method is discussed.
An inspiring computational work was done by Saha \textit{et al.}~\cite{saha10}.
The authors study a true atomistic model of a quantum-interference-controlled
molecular transistor formed by an 18-annulene attached between zigzag graphene nanoribbons.
The device was studied by a multi-terminal NEGF-DFT formalism~\cite{saha09}
and the current-switching effect was confirmed.

On the experimental side, ring-shaped tunnel nanojunctions can be formed by
cyclic organic molecules attached between two electrodes
(see for example~\cite{Kiguchi08,Kiguchi12,Bai12,Hong12}).
Conductors can be formed by metallic (usually Au or Pt) electrodes or by graphene
nanoribbons~\cite{nanoribbon_strips,nanoribbon_FET}.
These works employ direct attachment of particular molecules in between metallic
electrodes (Pt-C or Au-C bonds) using mechanically controllable break junctions.
Such contacts are reported to be more conductive and stable
than previously more common junctions employing a bridging thiol group
(thus forming a metal-S-C bond)~\cite{Reed1997}.
Another realisation of nanometer-scale sized rings is fabrication of structures
on proper substrates~\cite{Huant06,Huant07,Wieck10}.
This can be done using molecular-beam epitaxy, wet etching and optical or electron-beam lithography.
Finally we note that the concept of a quantum interference driven transistor
started to be more explicitly discussed in literature relatively recently~\cite{Cardamone06}.
It still represents an experimental challenge
and a major bunch of experimental results is yet expected to come.

The paper is organised as follows.
In Sec.~\ref{sec:model} we describe the model of the atomic ring.
In Sec.~\ref{sec:strobo} we explain the implementation of the stroboscopic wavepacket
method.
In Sec.~\ref{sec:current} we provide the formula
which we implement for electron current calculations.
Sec.~\ref{sec:eigenstates} describes stationary results as a basis from which
we move into non-stationary regime in section~\ref{sec:simul}, which contains
our main results.
%
%
\section{\label{sec:model} The model of the ring with contacts}
\begin{figure}[!t]
\centerline{\includegraphics[width=86mm]{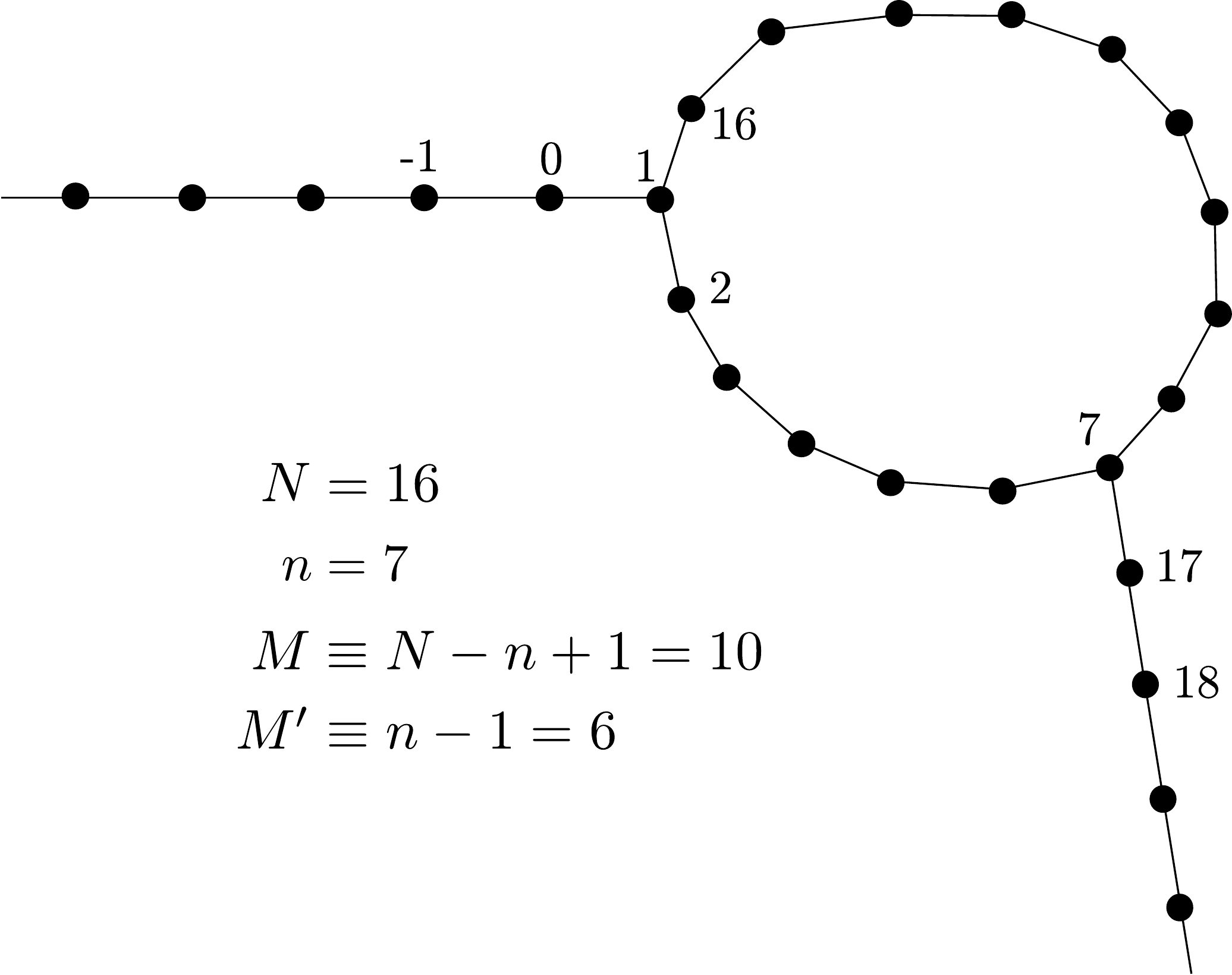}}
\caption{Topology of the studied systems and the numbering of atoms.
Throughout the paper we use $N$ for the total number of the ring atoms
as well as for the index of the atom just above the left vertex.
We use $n$ to denote the index of the right vertex atom.
Symbols $M$ and $M'$ are provided for convenience
to display relation to the notation of Ref.~\cite{Stefanucci09}.
}
\label{fig:ring}
\end{figure}

%
%
We consider a linear chain of atoms, each two being a lattice constant $a$ apart, 
with one finite ring which presents an obstacle for the flow of electrons
(Fig.~\ref{fig:ring}).
All couplings are considered within the TB approximation
in which we limit our treatment to one orbital per atom 
The chain is periodic apart from the ring region.
Hence, the TB Hamiltonian is of the form
\begin{equation}
\hat{H}(t) = \hat{H}^{0} + \hat{H}^{1} + \hat{H}^2(t)
\label{eq:hamilton_full}
\end{equation}
where
\begin{equation}
\hat{H}^{0}
=
\sum_{l=-\infty}^\infty \epsilon \; a^\dag_l a_l
+
\sum_{l=-\infty}^\infty \tB (a^\dag_{l+1} a_l +  a^\dag_l a_{l+1})
\label{eq:hamilton_TB}
\ .
\end{equation}
$a^\dag_l$ and $a_l$ are the fermionic creation and destruction operators,
$\epsilon$ is a constant on-site energy in the chain,
and $\tB$ is the TB hopping parameter here assumed to be negative.
The important simplifying assumption is that there is only one localised
atomic orbital (state $|l\rangle$) per each site.
The $\hat{H}^{0}$ represents the unperturbed chain, later referred to as the lead's Hamiltonian.

$\hat{H}^{1}$ is a stationary, periodicity-breaking term.
Its form corresponds to a TB ring structure, indicated in Fig.~\ref{fig:ring}.
(Specified values of $N$ and $n$ are taken only as an example there.)
%
%
In the simplest case, the modifications to the periodic chain Hamiltonian $\hat{H}^{0}$
due to the ring presence will be expressed
by the following matrix elements of $\hat{H}^{1}$:
\begin{equation}
\begin{array}{lllll}
H^1_{N,N+1} &=& H^1_{N+1,N} &=& -\tB
\\
H^1_{1,N} &=& H^1_{N,1} &=& \tB
\\
H^1_{n,N+1} &=& H^1_{N+1,n} &=& \tB
\end{array}
\ .
\label{eq:H1-topolog_pert}
\end{equation}
All other matrix elements of $\hat{H}^{1}$ are zeros.

Applied bias will be modelled by time-dependent shifts
of the on-site energies.
If a bias voltage $U(t)$ is applied then the on-site energies of the left lead
will take values $\epsilon + e U(t)$,
$e$ being magnitude of electron charge.
The operator $\hat{H}^2(t)$ expressed in the atomic-orbital basis
has the following matrix elements:
\begin{equation}
H^2_{l,l'}(t)
=
\left\{
\begin{array}{ll}
e U(t) \; \delta_{l,l'}  \ ,            & l \le 0 \ \textrm{(in the left lead)} \\
\frac{1}{2} e U(t) \; \delta_{l,l'} \ , & 1 \le l \le N \ \textrm{(within the ring).} \\
0 \ ,                                 & \textrm{all other} \ l, l'
\end{array}
\right.
\label{eq:H2}
\end{equation}
The on-site energies within the ring, if not differently specified, will thus
take values $\epsilon + e U(t)/2$.
The on-site energies in the right lead will remain unchanged, equal
to the equilibrium value $\epsilon$.

Three exceptions from the above prescription of the perturbation
$\hat{V}(t) = \hat{H}^{1} + \hat{H}^{2}(t)$ are studied:
(i)~In Sec.~\ref{sssec:slope}, where a uniform slope of the potential
energy within the ring is used instead of the spatially constant
value $e U(t)/2$,
(ii)~in Sec.~\ref{ssec:Vgate}, where a time- and branch- dependent
gate potential is is used,
(iii)~finally consideration of variable coupling strength
between the ring and the leads in Sec.~\ref{ssec:coupling}.
%
%
\section{\label{sec:strobo} The stroboscopic basis set}
\begin{figure}[t]
\centerline{\includegraphics[width=86mm]{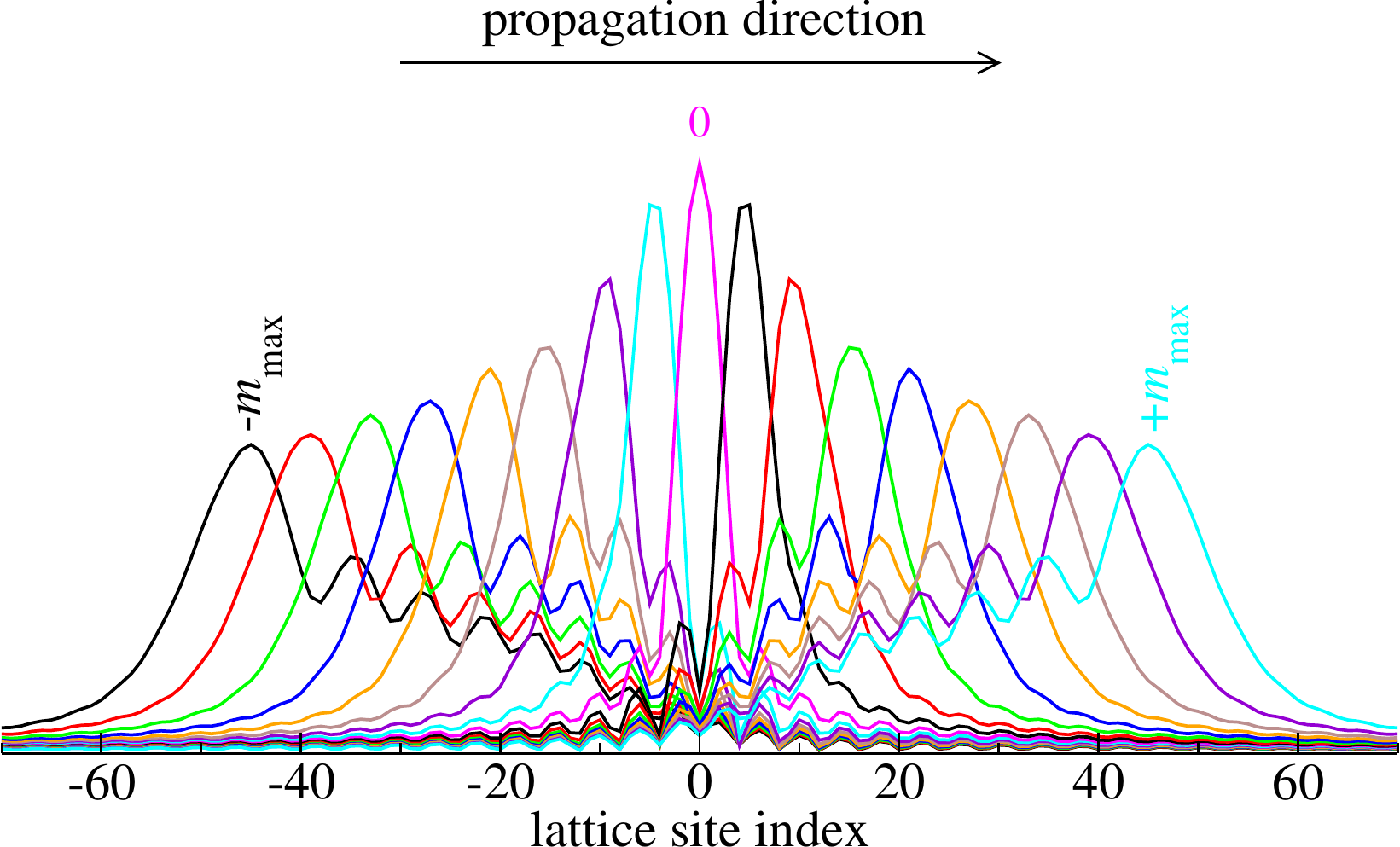}}
\caption{Schematic view on stroboscopic wave packets
$|n,\alpha,m;t\rangle$ at time $t=0$.
Actual picture shows absolute values of their projections to atomic orbitals $|l\rangle$,
i.e. the quantities $|\langle l|n,\alpha,m;t\rangle|$ as functions of 
the lattice site $l$.
Band index is chosen to be $n=1$, propagation direction
$\alpha=1$ and the $m$ indices run through the range $-8, -7, \dots, +8$.
The very low value of $\mmax=8$ is chosen for convenient visualisation
and is actual only for this scheme.
}
\label{fig:packets}
\end{figure}

%
%
The stroboscopic wavepacket basis set consists of wavepackets (Fig.~\ref{fig:packets}) constructed 
from the eigenstates of the unperturbed (lead's) Hamiltonian~(\ref{eq:hamilton_TB}).
The mathematical expression for the basis set vectors is~\cite{bokes_PRL,bokes_PCCP}
\begin{equation}
|n,\alpha,m;t\rangle
=
\exp\left[-\frac{i}{\hbar} (m \tau_n + t) \hat{H}^0\right]
\frac{1}{\sqrt{\Delta\mathcal{E}_n}}
\int_{\mathcal{E}_{n-1}}^{\mathcal{E}_n} \hspace*{-12pt} \dd\mathcal{E} \, 
|\mathcal{E},\alpha\rangle
\label{eq:basis_vec}
\end{equation}
where $\hat{H}^0$ is the lead's Hamiltonian and $|\mathcal{E},\alpha\rangle$ are its 
eigenstates normalised so that $\langle\mathcal{E},\alpha|\mathcal{E}',\alpha'\rangle
= \delta(\mathcal{E}-\mathcal{E}') \, \delta_{\alpha,\alpha'}$. 
Each basis function or wavepacket~(\ref{eq:basis_vec}) is uniquely characterised by three indexes: 
the band index $n$, the time shift index $m$, the degeneracy index $\alpha$ and time $t$.

\textbf{The band index} $\boldsymbol{n}$. 
The unperturbed Hamiltonian~(\ref{eq:hamilton_TB}) gives the dispersion relation 
\begin{equation}
E(k) = \epsilon + 2\tB \cos(k a) \ .
\label{eq:disprel}
\end{equation}
The corresponding energy range $[\epsilon + 2\tB, \epsilon - 2\tB]$
is in the SWPA divided into non-overlapping bands~\cite{bokes_PRL}.
$n^\mathrm{th}$ band has its energies $\mathcal{E} \in [\mathcal{E}_{n-1}, \mathcal{E}_n]$.
In our present work we use two equally wide bands ($\Nb = 2$)~\cite{bands}
spanning the whole TB energy range. The band index $n$ then takes values 1 a 2.

\textbf{The time-shift index} $\boldsymbol{m}$.
Within each band, different, mutually orthogonal basis functions are obtained by time
shifts $m \tau_n$, where the time step
\begin{equation}
\tau_n = \frac{2\pi\hbar}{\Delta\mathcal{E}_n}
\label{eq:tau_n}
\end{equation}
is set by the energy width of the band $\Delta\mathcal{E}_n = \mathcal{E}_n - \mathcal{E}_{n-1}$.
$m$ attains all integer values, but in numerical calculations it is
restricted to $m = -\mmax, \dots, +\mmax$, as it is discussed  at the end of this section.

\textbf{The degeneracy index} $\boldsymbol{\alpha}$.
Apart from the energy, each eigenstate of the lead's Hamiltonian has further 
quantum numbers. In the present system, the only one is the direction of propagation 
of the Bloch eigenstates: those propagating from the left
to the right (index $\alpha = +1$) and the opposite ones (index $\alpha = -1$).

\textbf{The time} $\boldsymbol{t}$.
The time $t$ in the notation indicates that the basis state is unitarily 
propagated~\cite{bokes_PCCP} by the lead's Hamiltonian~(\ref{eq:hamilton_TB}).
The use of this ``moving'' basis set significantly simplifies the form of the SchE
in the SWB representations (see Eq.~\ref{eq:diffeq} below), very much in the spirit of using the 
interaction representation in formal perturbation theory in quantum theory.
At each given time $t$ the SWB vectors form an orthonormal system since
\begin{equation}
\langle n,\alpha,m;t|n',\alpha',m';t\rangle
=
\delta_{n,n'} \delta_{\alpha,\alpha'} \delta_{m,m'}
\ .
\end{equation}
The set of basis states~(\ref{eq:basis_vec}) is complete if $\mmax$ is infinite.
%

Given the form of Hamiltonian~(\ref{eq:hamilton_full}),
each electron in the system evolves independently as described by the
SchE
\begin{equation}
i \hbar \frac{\partial}{\partial t} |\Psi(t)\rangle
=
\hat{H}(t) |\Psi(t)\rangle
\ .
\label{eq:SchE}
\end{equation}
State vectors $|\Psi(t)\rangle$ for each electron are expanded in
this basis set:
\begin{align}
\nonumber
|\Psi(t)\rangle
& =
\sum_{n=1}^{\Nb} \sum_{\alpha=\pm 1} \sum_{m=-\mmax}^{\mmax} A_{n,\alpha,m}(t) \; |n,\alpha,m; t\rangle
\equiv
\\
& \equiv \sum_{o=1}^{2 \Nb (2\mmax+1)} A_o(t) |o;t\rangle
\ ,
\end{align}
where we have introduced a composite index $o = (n,\alpha,m)$.
The SchE provides a linear system of differential equations
for each of the involved electrons.
Due to the time-evolution of the basis functions given by the operator $\hat{H}^0$,
the equations for amplitudes $A_o(t)$ include matrix elements of the perturbation only:
\begin{equation}
i \hbar \frac{\dd A_o(t)}{\dd t}
=
\sum_{o'=1}^{2 \Nb (2\mmax+1)} 
\langle o;t|\hat{V}(t)| o';t\rangle \; A_{o'}(t)
\label{eq:diffeq}
\ ,
\end{equation}
with $\hat{V}(t) = \hat{H}^{1} + \hat{H}^2(t)$.
This is of key importance for our method: basis set wavepackets, which 
are localised far from the central region feel either zero perturbation
(in the right lead, and hence their matrix elements are zeros)
or they feel only the spatially uniform bias-induced potential of the left lead;
see eq.~(\ref{eq:H2}).
For such matrix elements we have
\begin{align}
\label{eq:bias_matel}
&\langle o;t|\hat{V}(t)| o';t\rangle
\approx
e U(t) \delta_{o,o'}
\\
\nonumber
&\textrm{in the left lead, for packets far from the central region}
\end{align}
meaning that probability amplitudes of the wavepackets evolve freely,
(independently of other amplitudes) far in the left lead.
Situation is even simpler far in the right lead where $\hat{V}(t)$ is zero
and corresponding amplitudes do not evolve at all.
The constant or freely evolving amplitudes need not be explicitly included
into simulation and this omission in principle does not have any impact on 
accuracy of the simulation.
In practice, since the SWPs are not strictly localised,
by using the finite cutoff $\mmax$ we introduce certain error into simulations.

The simple form of equations of motion~(\ref{eq:diffeq}) has been possible
due to the employment of the moving (i.e. unitarily propagating) basis set.
The simplicity is in that the matrix elements are computed only from the
interaction term $\hat{V}(t) = \hat{H}^{1} + \hat{H}^2(t)$
of the total Hamiltonian~(\ref{eq:hamilton_full}).

Wavepackets corresponding to vectors~(\ref{eq:basis_vec}) unitarily propagate
in time which implies that each particular packet
will travel away from the central region after some time.
The result would be
that (for \emph{finite} $\mmax$) at large times the region would not be covered by basis set
wave packets at all and no electrons would be presents in the central region at long times.
In detail, the unitary propagation of the wavepackets is such that
during a time interval $\tau_n$ [defined below~(\ref{eq:basis_vec})] each packet
moves exactly to the position of the neighbouring packet
(either on its left or on its right, depending on the propagation direction).
If we had a very large basis set ($\mmax \to \infty$) and were
looking at a movie of the wavepackets only at the ``stroboscopic'' times
$m' \tau_n$,
we could not see any motion of the packets.
For any finite $\mmax$ we could observe similar static picture only for a limited
time and in a limited spatial region because the wavepackets constantly
propagate
from one's position to the other's.

To prevent the gradual disappearance of the basis functions from the central region
in simulations, we periodically insert new basis wave packets into
the system at every period $\tau_n$.
The newly inserted packets are localised far away from the centre, but moving toward it.
To avoid the increase of the total number of basis vectors,
we also periodically remove
basis wave packets (those which has left the central region).
In other words, we remove the leading wave packet from the train of 
the propagating packets and attach a new wave packet at the tail of the train.
(See also Fig.~\ref{fig:packets}.)
This removal/insertion procedure is accomplished for each band (indexed by $n$)
independently; different bands could in principle have different widths
and consequently different parameters $\tau_n$.
For given band $n$ there are two such procedures: one for $\alpha=-1$
(packets propagating from right to left), the other for $\alpha=+1$
(packets propagating from left to right).
In this way we keep the basis set vectors in the region of interest
(the ``central region'') and also keep the number of the vectors constant.

It is necessary to have $\mmax$ sufficiently large so that the basis vectors
$|n,\alpha,m;t\rangle$ with $m \approx \mmax$
are decoupled from other basis vectors
(see more reasoning below) and hence the removal
of the basis vector $|n,\alpha,\mmax;t\rangle$ does not affect the quantum
dynamics of the electron, amplitude of which has been removed.

In addition to the basis vector removal/insertion procedure,
we also track individual electrons in the simulation.
Each newly inserted basis state,
if belonging to normally occupied band ($n=1$ in this work), is set to
be occupied at the instance of the insertion.
Also, an electron, which through time evolution gets far from the central
region, is removed from the simulation according to a proper criterion.
Therefore the simulation explicitly describes an \emph{open} system in which the
total number of electrons generally fluctuates~\cite{explain_ElRemove}.

The initial conditions applied to the probability amplitudes are such that
\begin{equation}
A_{n,\alpha,m}(0) = 
\left\{
\begin{array}{ll}
1, \ \ \textrm{for}\ n = 1  \\
0, \ \ \textrm{for}\ n = 2
\end{array}
\right.
\ , \ \ \ \textrm{for all} \ \alpha, m,
\label{eq:init_state}
\end{equation}
i.e.  only the lower band is initially occupied.

Matrix elements of $\hat{V}(t)$ are evaluated using the overlaps
$\langle l|o;t\rangle$.
In Appendix we derive that the overlaps can be expressed by the formula
\begin{align}
\nonumber
& \langle l | n, \alpha, m; t \rangle
\equiv 
\langle l | o; t \rangle
=\\
\nonumber
&=
\sqrt{\frac{|\tB|}{\pi \Delta\mathcal{E}_n}}
\exp\left[-\frac{i}{\hbar} \epsilon \left(m \tau_n + t\right)\right]
\\
&\times\int_{\mathcal{K}_{n-1}}^{\mathcal{K}_n}
\hspace*{-12pt} \sqrt{\sin\mathcal{K}} \;
\exp\left\{
i \left[\alpha \mathcal{K} l - 2 \frac{\tB}{\hbar} \left(m \tau_n + t\right) \cos\mathcal{K}\right]
\right\}
\, \dd\mathcal{K}
\ .
\label{eq:overlap}
\end{align}
Quantities $\mathcal{K}_n \equiv k_n a$ are dimensionless wavenumbers
corresponding to energies $\mathcal{E}_n$ through the TB dispersion relation~(\ref{eq:disprel}).
The cutoff $\lmax$ on atomic sites indexed by $l$ has to be chosen
sufficiently large
in order to cover all SWPs~(\ref{eq:basis_vec}) included into simulation.
For majority of calculations we use $\mmax = 352$ with rather conservatively
chosen $\lmax=3016$.
In one case we use $\mmax=1200$ and corresponding $\lmax=7576$.
Integrals in~(\ref{eq:overlap}) have to be evaluated numerically.

The SWPA as described above and used in this work does not include any electron-electron (e-e)
interactions except for keeping them always in strictly orthogonal states.
We expect that e-e interactions with effects like Coulomb blockade among others
become more important for weakly coupled rings (although they can never be neglected in any
accurate quantitative description)~\cite{e-e}.
Further developments of the SWPA will include also realistic e-e interactions
at least on a mean-field level of description like in time-dependent Hartree-Fock
theory or in TDDFT.
Presently we could only include a simple dynamical mean-field interaction according to a chosen model.
Having done this we have not found any significant interesting impacts of the interaction.
Hence we decided not to include any such results in the present work
as it would present an unnecessary complication.

The system of the equations~(\ref{eq:diffeq})
is solved using
the modified-midpoint method~\cite{numrec}.
Its accuracy and stability is fully sufficient for given problem.
As a complementary treatment applicable for stationary currents, we
compute exact stationary currents for given
TB Hamiltonian~(\ref{eq:hamilton_full})
in cases when it is time-independent.
See Sec.~\ref{sec:eigenstates} for more details.
%
%
\section{\label{sec:current}Electron current formula}
%
In this section we briefly provide an expression for local electron current.
For this purpose we 
consider one-dimensional infinite TB chain, specified in Sec.~\ref{sec:model}.
For such a system the local current operator at bond between sites $l_0$ and  $l_0+1$
is given by the formula (see Ref.~\cite{NEGF}, p.~162 therein).
\begin{align}
&\hat{I}_{s,l_0}
=
\frac{1}{i\hbar} \left(
H_{l_0+1,l_0} \; a^\dag_{s,l_0+1} a_{s,l_0}
-
H_{l_0,l_0+1} \; a^\dag_{s,l_0} a_{s,l_0+1}
\right)
\label{eq:loc_curr_operator}
\\
\nonumber
&\textrm{with} \ \ s = \pm \frac{1}{2}
\end{align}
being the spin index.
Taking the expectation value in a single-electron state $|\Psi(t)\rangle$
we obtain
\begin{equation}
I^{(1)}_{s,l_0}(t)
=
\frac{1}{i\hbar} H_{l_0+1,l_0}(t) \;
\langle \Psi(t) | l_0+1 \rangle \; \langle l_0 | \Psi(t) \rangle
+
\textrm{c.c.}
\label{eq:current_1ele}
\ .
\end{equation}
Superscript $(1)$ marks that it is a current caused by single electron;
we must sum up over all electrons in the system to obtain the total current.
If we express the state vector in the stroboscopic basis orbitals then we
obtain
\begin{align}
\nonumber
I^{(1)}_{l_0}(t)
& =
\frac{2}{i\hbar} H_{l_0+1,l_0}(t)
\left[\sum_{o=1}^\infty A_{o}^*(t) \langle o; t| l_0+1 \rangle\right] \\
&\times \left[\sum_{o'=1}^\infty A_{o'}(t) \langle l_0 | o'; t\rangle\right]
+
\textrm{c.c.}
\label{eq:current_1ele_strobo}
\ .
\end{align}
The factor of 2 has been added to take into account two electrons differing
only by their spins.
As for the sign convention used for the current
expressed by eqs.~(\ref{eq:current_1ele}), (\ref{eq:current_1ele_strobo}),
it indicates flow of particles (electrons), not the charge.
The same purely local result for $I^{(1)}_{l_0}(t)$
as shown above could be derived also for currents in particular ring arms.
Hence we use formula~(\ref{eq:current_1ele_strobo}) to compute time-dependent current
(contribution from one electron)
through any chain (lead or arm of the ring) of the complete system.
The total current is obtained by summing up contributions~(\ref{eq:current_1ele_strobo})
generated by all individual explicitly included electrons in the system:
\begin{equation}
I_{l_0}(t) = \sum_\mathrm{ele=1}^{N_{\mathrm{ele}}} I^{(\mathrm{ele})}_{l_0}(t)
\label{eq:current_total_strobo}
\end{equation}
%
%
\section{\label{sec:eigenstates}Stationary currents through ring
nanojunctions}
%
%
Currents through TB rings have been studied in literature mostly in stationary regimes.
It is convenient to compare quasi-stationary currents from our
method to exact stationary results for the TB model.
For this purpose,
the whole system -- ring with leads -- is assumed to be composed of identical 
atoms described by the simple TB Hamiltonian with all couplings equal to $\tB$.
The only perturbation to periodic chain Hamiltonian $\hat{H}^0$ are then
terms~(\ref{eq:H1-topolog_pert}) which represent a perturbation to the
topology of the linear chain.
It was shown in Ref.~\cite{Stefanucci09} that in the stationary regime
there are both conducting and insulating configurations depending on where
the leads are attached to the ring.
Specifically, all odd-numbered rings are always conductive irrespective of
the choice of attachment site.
On the other hand, even-numbered rings can exhibit both behaviours.
These findings are confirmed and extended by the exact TB results 
of Ref.~\cite{SGML11}.

We have obtained stationary results from 
exact eigenstates of the TB Hamiltonian
$\hat{H}^0 + \hat{H}^1 + \hat{H}^2$ defined in Sec.~\ref{sec:model}.
This approach is, for the Hamiltonian used in our work,
equivalent to the method described
in Ref.~\cite{SGML11}.
As specified in Sec.~\ref{sec:model}, the
applied bias $U$ is modelled by lifting on-site energies in the left 
(source) lead at the value $\eL = \epsilon + e U$.
The on-site energies in the right lead (drain) are kept at 
$\eR = \epsilon$.
On-site energies of all ring atoms are kept at the intermediate value
$\eS = \epsilon + e U/2 = (\eL + \eR)/2$.
TB couplings are unchanged, i.e. they all are equal to $\tB$.
The eigenstates for such system
are expressed in the TB basis,
\begin{equation}
|\psi\rangle  =  \sum_{l=-\infty}^\infty  \psi_l |l\rangle
\ .
\end{equation}
They depend on the eigenenergy $E$ and the direction of propagation $\alpha$.
For brevity we do not show these dependences explicitly.
We compactly represent the amplitudes by composite formula
\ifthenelse{\boolean{twocol}}{%
\begin{align}
\nonumber
&\psi_l
=\\
\nonumber
\\
&\left[\ \ \ \;
\renewcommand{\arraystretch}{1.8}
\arraycolsep-12pt
\begin{array}{lll}
                                                    &   \mathcal{P} \; f(\KS l) + \mathcal{Q} \; f(-\KS l)  &                                 \\
\mathcal{A} \; f(\KL l) + \mathcal{B} \; f(-\KL l)  &                                                       &   \ \ \ \mathcal{C} \; f(\KR l) \\
                                                    &   \mathcal{F} \; f(\KS l) + \mathcal{G} \; f(-\KS l)  &
\end{array}
\renewcommand{\arraystretch}{1.0}
\ \ \ \; \right]
\label{eq:psi_new_aligned}
\end{align}
}
{%
\begin{equation}
\psi_l
=
\left[
\begin{array}{lll}
                                                    &   \mathcal{P} \; f(\KS l) + \mathcal{Q} \; f(-\KS l)  &                           \\
\mathcal{A} \; f(\KL l) + \mathcal{B} \; f(-\KL l)  &                                                       &   \mathcal{C} \; f(\KR l) \\
                                                    &   \mathcal{F} \; f(\KS l) + \mathcal{G} \; f(-\KS l)  &
\end{array}
\right]
\label{eq:psi_new_aligned}
\end{equation}
}
with
\begin{equation}
f(\mathcal{K} l)
=
\sqrt{\frac{\rho(E)}{2\pi}} e^{i \mathcal{K} l}
\label{eq:weighted_exp}
\ .
\end{equation}
The specific form of~(\ref{eq:psi_new_aligned}) is an abbreviated non-standard notation
mimicking the spatial positions of particular system chains
and has to be understood in the sense that
$\psi_l = \mathcal{A} \; f(\KL l) + \mathcal{B} \; f(-\KL l)$ for $l \in \ $ left lead,
$\psi_l = \mathcal{F} \; f(\KS l) + \mathcal{G} \; f(-\KS l)$ for $l \in \ $ lower ring branch, etc.
We introduced the dimensionless k-number
\begin{equation}
\mathcal{K} \equiv k a
\end{equation}
with $a$ being the lattice constant.
$\rho(E)$ is the local density of states (DOS) of an ideal infinite chain
for a particular branch of the system and for the TB model is expressed as
\begin{equation}
\rho(E)
=
\frac{1}{
\sqrt{4 \tB^2 - (E-\epsilon)^2}
}
\ .
\end{equation}
$\epsilon$, identified with $\eL$, $\eS$ or $\eR$,
is the value of the on-site energies in given branch of the system.
In the model under study it also represents Fermi energies of
particular bulk systems.
Indices L, S and R stand for the left lead, small system (the ring)
and the right lead, respectively~\cite{explain_KL}.
%
\begin{figure}[t]
\centerline{\includegraphics[width=86mm]{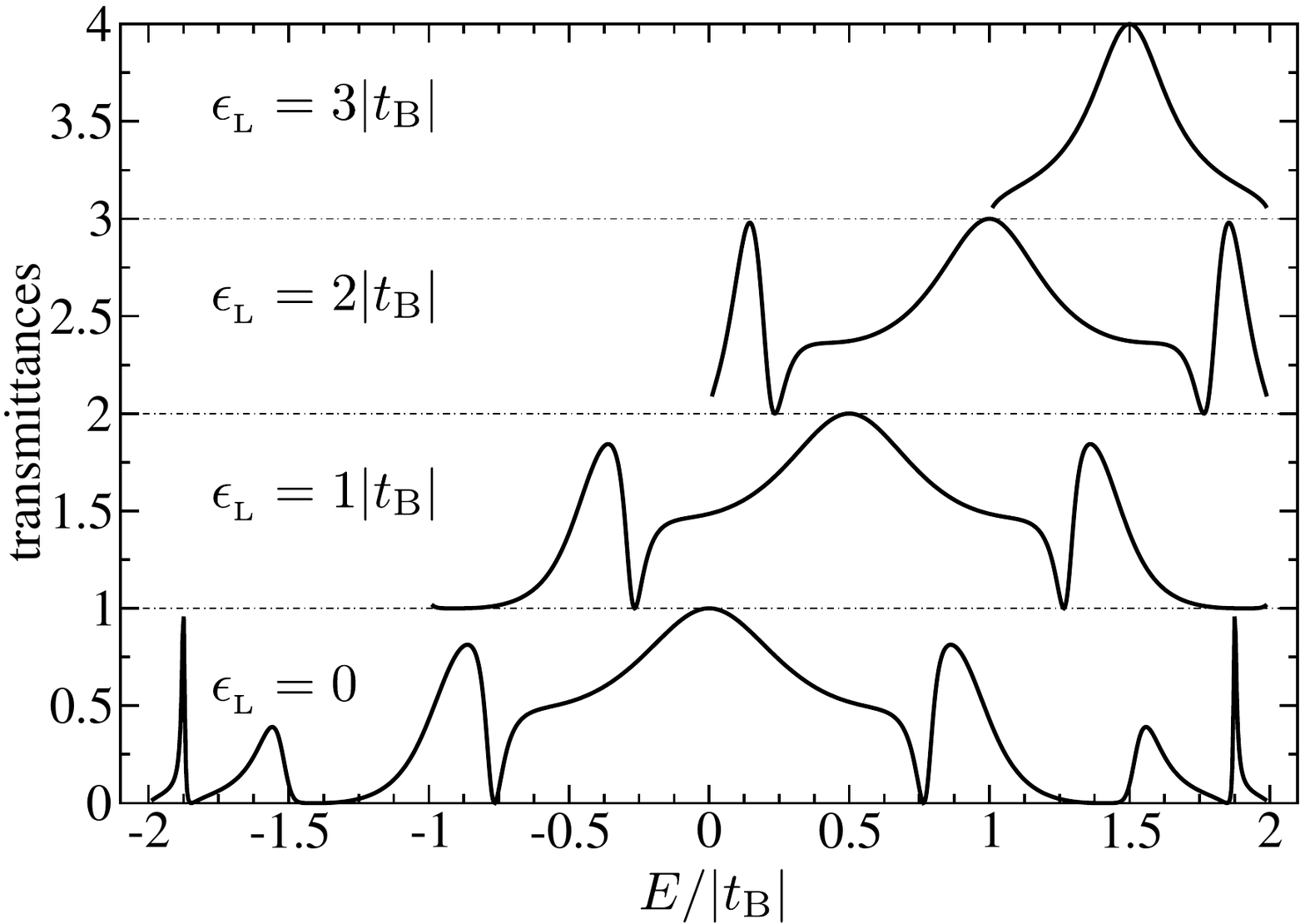}}
\caption{Transmittances of the ring depicted in Fig.~\ref{fig:ring}.
($N = 16$, $n = 7$ or equivalently $n = 11$)
calculated from exact eigenstates of the TB Hamiltonian;
see text and Ref.~\cite{SGML11}.
The plots with higher $\eL$ values are vertically shifted for 
convenience and the shifts are visually enhanced by horizontal
dashed lines.
Parameters $\eL$ are on-site energies in the left lead.
Right leads has always set $\eR = 0$.
The small system (ring) has $\eS = (\eL + \eR)/2$.
The transmittances are computed and shown only for energies at which 
extended propagating eigenstates exist.
$\tB < 0$ is the tight-binding hopping parameter, magnitude of which is
used as the unit of energy throughout this work.}
\label{fig:transmittances}
\end{figure}
%
When convenient we use also notations $\rho_\mathrm{L}$, $\rho_\mathrm{R}$ and $\rho_\mathrm{S}$ to 
distinguish DOSs in the left lead, right lead and in the small system (ring).
Parameters $\mathcal{A}$ and $\mathcal{B}$ describe wavefunction of the left
lead, parameter $\mathcal{C}$ in the right lead,
$\mathcal{F}$ and $\mathcal{G}$ in the lower ring arm
and finally $\mathcal{P}$ and $\mathcal{Q}$ in the upper ring arm.

An electron being in particular eigenstate $|\psi\rangle$ is then interpreted
as partially reflected and partially transmitted, with transmittance 
computed as
\begin{equation}
\mathbf{T}(E)
=
\left|\frac{\mathcal{C}}{\mathcal{A}}\right|^2
\ .
\label{eq:T_C_A}
\end{equation}
This simple formula is applicable also in cases when the Fermi levels
in the two leads are different.
In such a case we need additional factors involving the respective
group velocities.
These factors have been explicitly inserted into functions~(\ref{eq:weighted_exp})
hence are not explicitly present in the formula~(\ref{eq:T_C_A}).
The results from exact eigenstates for the ring of $N=16$ sites and $n=7$
are shown on Fig.~\ref{fig:transmittances}.

Having transmittances available, we compute the stationary electron currents 
in the leads using the Landauer formula
\begin{equation}
I = \frac{2 e}{h} \int_{\EF}^{\EF+eU} \mathbf{T}(E) \, \dd E
\ .
\label{eq:Landauer}
\end{equation}
We remind that the transmittance $\mathbf{T}(E)$ depends parametrically
also on the chemical potentials in the leads (which are equal to 
$\eL$ and $\eR$) and on the on-site energies of the ring atoms
which are all equal to $\eS$.
In the eigenstate approach we always use $\eR = 0$ and
$\eL = e U$ (applied bias) and the on-site energies of the ring are set to
$\eS = (\eL + \eR)/2$.

\begin{figure}[t]
\centerline{\includegraphics[width=86mm]{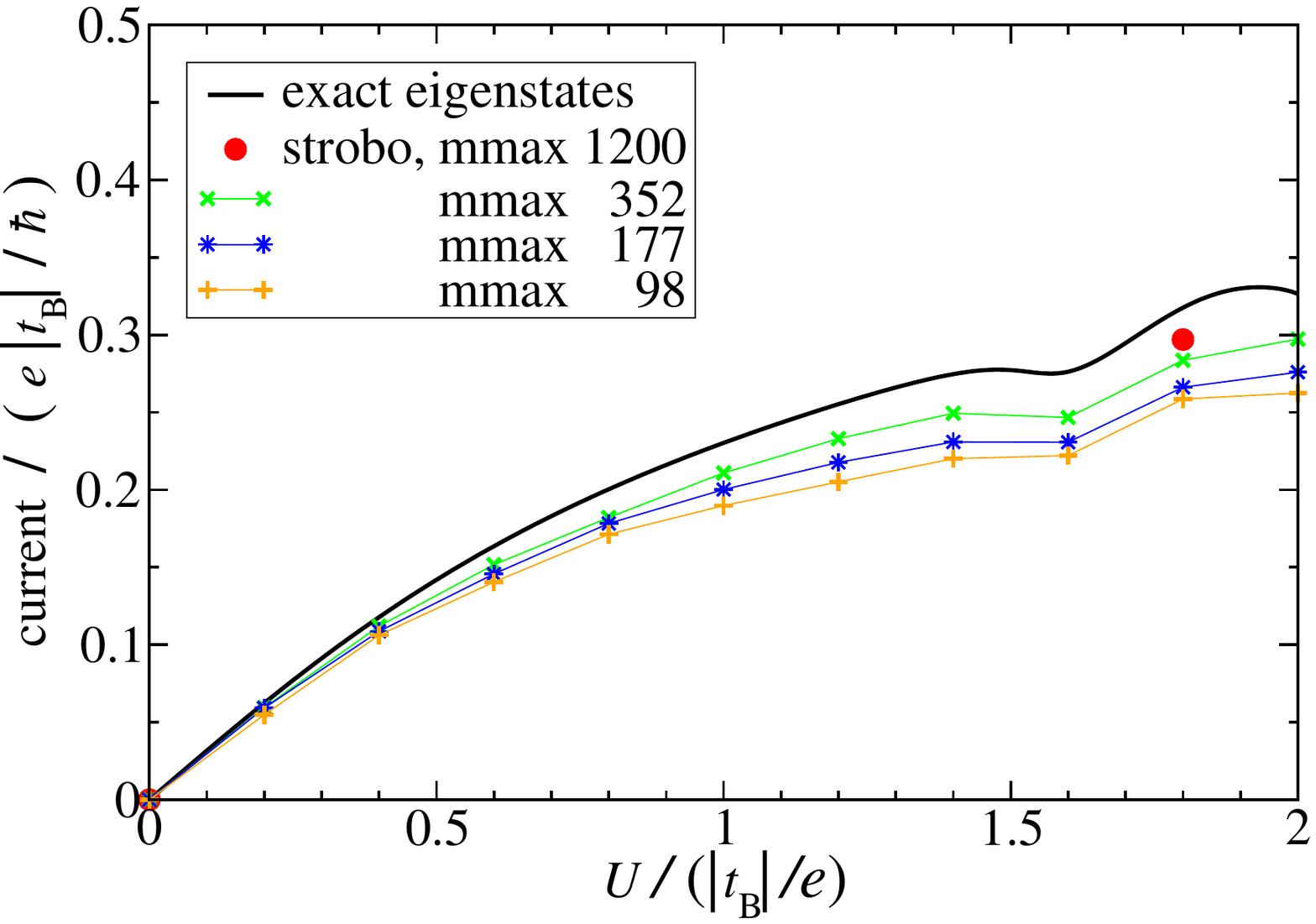}}
\caption{Stationary electron currents \textit{vs.} bias through the ring system
depicted on Fig.~\ref{fig:ring} ($N = 16$, $n=7$) calculated by 
the two different approaches.
Solid black line shows electron current obtained from
the Landauer formula~(\ref{eq:Landauer})
in which the transmittances,
shown on Fig.~\ref{fig:transmittances},
have been computed from 
the exact solution of the TB model.
Discrete symbols show long-time quasi-stationary values from our time-dependent
SWPA at several basis sizes given by the values
of $m_\mathrm{max}$~\cite{bokes_PRL,bokes_PCCP}.
In most of this work we use either $m_\mathrm{max} = 352$ or $m_\mathrm{max} = 1200$.
In these and all other numerical results throughout this work we consider
the periodic tight-binding
model of the leads with zero Fermi energy in equilibrium state.
Out of equilibrium, Fermi level in the left lead is lifted up to the bias
value $U$ while keeping the right lead at zero Fermi energy.
See main text, most importantly subsection~\ref{sssec:slope}, for further details.
}
\label{fig:stat_curr}
\end{figure}

The integral in Eq.~(\ref{eq:Landauer}) is done numerically.
In Fig.~\ref{fig:stat_curr} we show the comparison of the exact stationary currents
and the currents obtained
by the time-dependent wavepacket approach.
%
%
%
Results from the SWPA (discrete symbols)
represent long-time
quasi-stationary values from time-dependent calculations.
We see overall agreement between the two approaches,
especially for low biases.
The results from eigenstates are exact for given Hamiltonian.
The SWPA~\cite{bokes_PRL,bokes_PCCP} is in principle exact
too, but its accuracy is lowered by
the finite cutoff to the number of basis functions which is
$2 \Nb (2 \mmax+1)$.
The impact of the finite cutoff is typically not important at low biases
but increases with the bias and the obtained accuracy is then lower at high
bias conditions.
Convergence of the current with the cutoff becomes very slow at large $\mmax$.
Several examples computed with different basis set sizes are shown
on Fig.~\ref{fig:stat_curr}.
The steady-state current in the leads is always underestimated~\cite{I_in_ring}
when finite $\mmax$ is used.
To understand the finite basis set impact, we 
first remark that the dynamics of an individual electron becomes (quite obviously)
more accurately described when using larger $\mmax$,
hence providing more accurate current~(\ref{eq:current_1ele_strobo}) generated by single electron.
Second, the increased $\mmax$ results in a larger number of electrons
$N_{\mathrm{ele}}$ included
into the total current formula~(\ref{eq:current_total_strobo}).
As a consequence, the precise dependence of the current on
$\mmax$ is a complicated function.
The much slower convergence at higher values of $\mmax$ 
can qualitatively be understood as a result of larger spread of the SWPs
at hight $m$ (shown on Fig.~\ref{fig:packets}).
In the limit of $m \to \infty$ the packets become fully delocalised.
Individual high-$m$ packets contribute very weakly to the total local
current at given point in space.
The less converged results in absolute terms are pronounced especially at larger biases.
The convergence is worse also in relative terms:
at $U = 0.2\,|\tB|/e$ we reach $95.5\%$ of the exact limit (Fig.~\ref{fig:stat_curr}).
At $U = 0.5$ we get $91.6\%$ and at $U = 1.8$ only $89.3\%$, 
all at $\mmax = 352$.
The worse convergence at larger biases can be qualitatively understood
on the basis of the equations of motion~(\ref{eq:diffeq}) and bias-related
matrix elements~(\ref{eq:bias_matel}).
Using a finite $\mmax$ in~(\ref{eq:diffeq}) we drop some portion of the
exact equations of motion, in particular some of the terms~(\ref{eq:bias_matel})
which describe the interaction with bias-induced lifts of on-site energies in the left lead.
Using higher bias increases significance of those matrix elements and hence 
makes convergence more difficult.
The finite basis set size would not impact the accuracy of our approach if the stroboscopic
basis states with large $|m|$ indices were well localised in space.
In reality, even wavepackets with very large $|m|$ indices exhibit long tails
towards the central point of the lattice where the nanojunction is placed.
Ideally, having basis wavepackets (those with large $|m|$ indices)
with no tails in the nanojunction region
would result in perfectly converged results.
(Such an ideal situation is impossible because of the dispersion which
spreads the wavepackets.
Our further effort in development of the SWPA is directed to overcome
the convergence issues despite the presence of the dispersion.)
%
%
\section{\label{sec:simul}Simulations}
%
%
\subsection{Bias voltage switch on ring-shaped nanojunction}
%
%
Time-dependent bias is modelled by given time-dependent (but spatially
uniform) lift of the TB on-site energies in the left lead only.
The atoms in the right lead have their on-site energies all set to zero,
which is also the value of Fermi energy in equilibrium.
%
\begin{figure}[!t]
\centerline{\includegraphics[width=86mm]{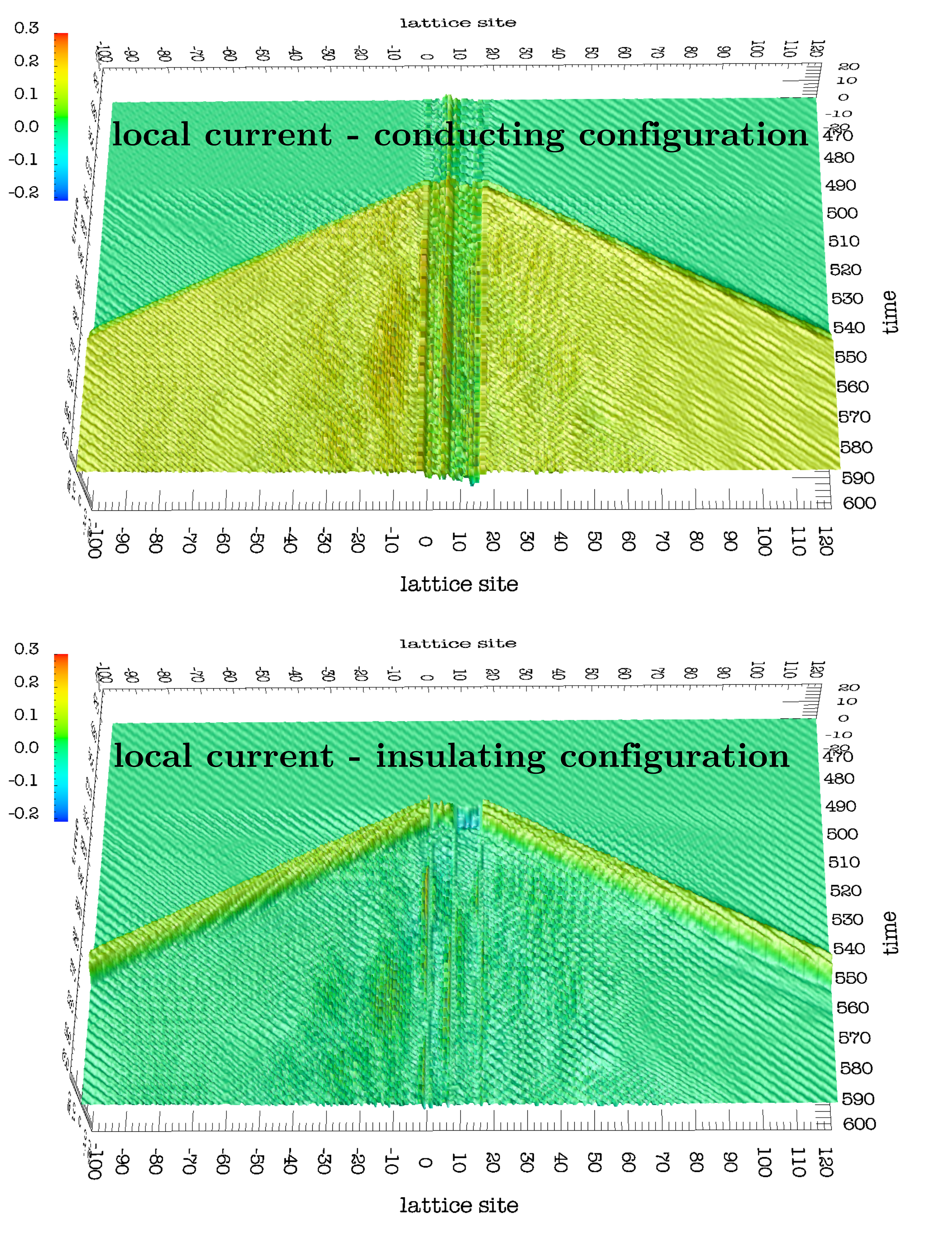}}
\caption{%
\textbf{Upper panel:} Local electron current as a function of the two
variables, lattice site and time,
computed for conductive ring configuration with $N = 16$
and $n = 7$.
Lattice sites from $1,\dots,16$ belong to the ring structure.
The applied bias of $0.5\,|\tB|/e$
was abruptly switched on at time $t_\mathrm{sw} = 500\,\hbar/|\tB|$.
Currents through vertex sites (here 1 and 7) do not have well defined values
and certain interpolated values have been used for visualisation purposes.
\textbf{Lower panel:} Analogous plot but now for insulating configuration with
$n = 8$.
Basis set size for these plots uses $m_\mathrm{max} = 352$.
Time is expressed in units of $\hbar/|\tB|$ and local electron currents
(on the $z$ axes) in units of $e|\tB|/\hbar$.
On this figure we do not apply any artificial smoothening of the data
(see caption to Fig.~\ref{fig:vertex_dep} for example).
}
\label{fig:CondInsu_SurfPlots}
\end{figure}
%
The atoms within the ring have their on-site energies set to the
half of the bias value applied at given time although we test also different
model in subsection~\ref{sssec:slope}.
In our simulations we first ``equilibrate'' given system by letting it evolve
at zero bias (and zero temperature) for time up to $\tsw=500\,\hbar/|\tB|$.
The ``equilibration'' evolves
the non-interacting many-electron initial state of the system,
described by eq.~(\ref{eq:init_state}),
into (again non-interacting) stationary many-electron state of the Hamiltonian with the localised perturbation
(the ring system and eventual variations of some of the on-site energies 
or interatomic couplings).
We have verified that the equilibration period is sufficient to obtain
converged transient electric currents in the sense that these currents
are practically independent of particular choice of $\tsw$
if $\tsw$ is at least $500\,\hbar/|\tB|$.
The transient electron currents are evaluated according to
formula~(\ref{eq:current_1ele_strobo}), with applied summation over all contributing
electrons in the system.
We study how these transient currents
depend on several ring parameters like the ring size, placement of the 
terminals and the magnitude of applied bias.
Typical behaviour of the transient electron currents is shown
in Fig.~\ref{fig:CondInsu_SurfPlots}.
It is noticeable that both the temporal and spatial evolution must be considered.
The intersect of the two ridges corresponds to the time when the bias was
turned on and to the spatial location of the ring.
The slope of the ridge corresponds to the Fermi velocity,
$v_\mathrm{F} = (1/\hbar) \, \partial_k E(k\!=\!\pi/a) = -2 a \tB/\hbar$,
so that the time interval between the instance of switching-on the bias 
and the arrival of the density or current  perturbation to a site $l$  
is given by the expression $t_l = l a/v_\mathrm{F}$.
The ridge on the left side corresponds to the reflected electron impulse
propagating back into the left lead.
Similarly, the other ridge represents the transmitted wavefront in the right lead.
The function values are mostly positive meaning that the
currents in both leads are positive (i.e. electrons flow from the source lead
to the drain lead).
Results for several specific cases are discussed in the following subsections.
Presented ring sizes $N$ vary from 16 to 18.
We have calculated also several other sizes, starting from $N=5$, not shown here.
Plotted local currents will typically correspond to a lattice site positioned
far from the ring, in most cases site $l = 120$, which is displaced about
100 lattice constants from the considered rings.
%
\subsubsection{Dependence on the location of drain vertex}
%
%
\begin{figure}[t]
\centerline{\includegraphics[width=86mm]{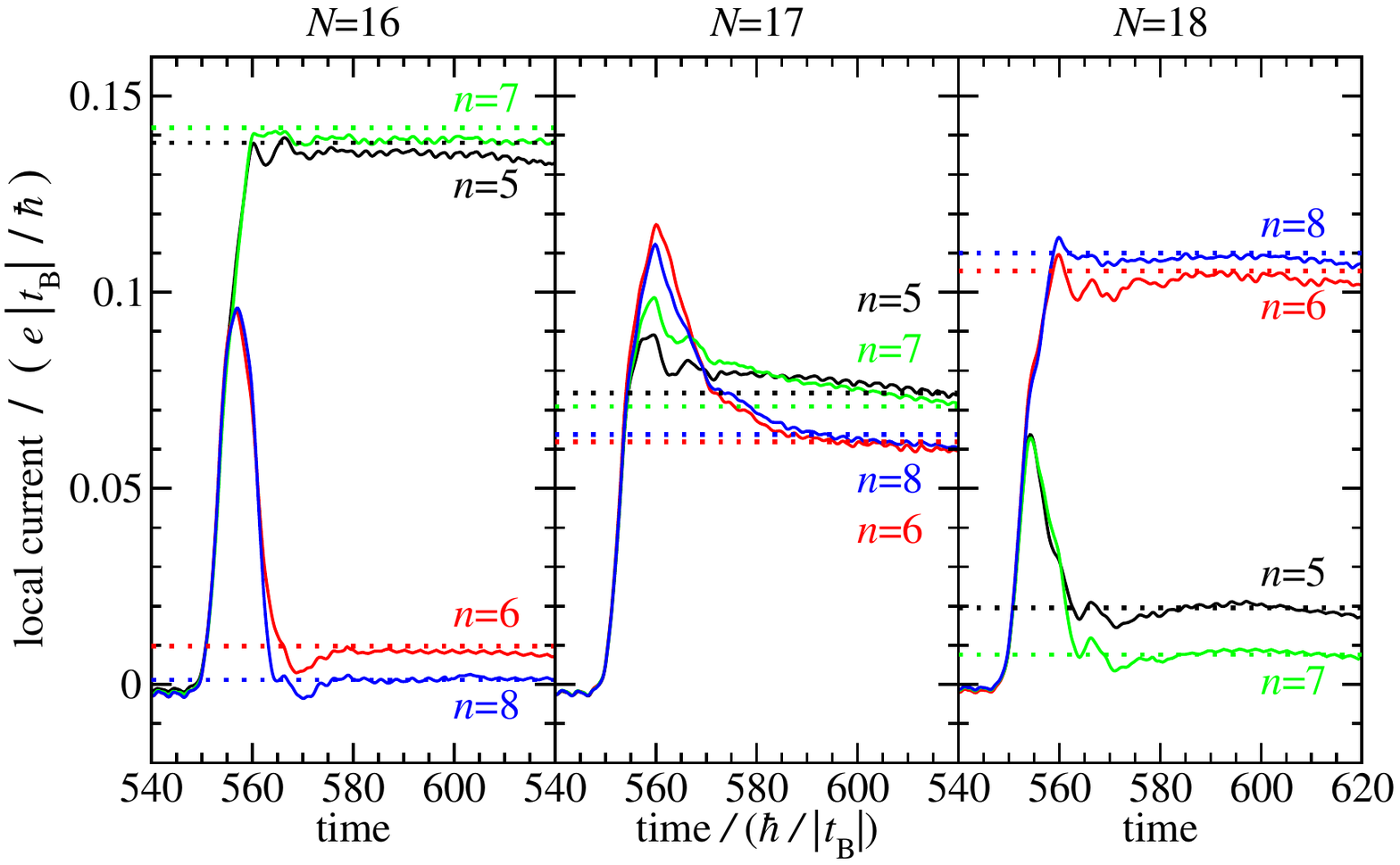}}
\caption{Transient currents computed in the drain lead
at site $l = 120$ for several different drain lead attachments
(vertices specified by the numbers $n = 5, 6, 7, 8$ shown at individual plots)
of the rings composed of $N = 16, 17$ and 18 atoms.
In all cases
the applied bias $U = 0.5\,|\tB|/e$ has been switched on abruptly at time
$t_\mathrm{sw} = 500\,\hbar/|\tB|$.
Basis set size again uses $\mmax = 352$.
The plots have been smoothed by taking running averages over an interval 
of about $\pi\,\hbar/|\tB|$.
The dashed lines represent analytically computed stationary currents as obtained
by the exact-eigenstate method described in Sec.~\ref{sec:eigenstates}.
}
\label{fig:vertex_dep}
\end{figure}
%
The dependence of the time-dependent current on  the position of the right (drain)
lead is shown in Fig.~\ref{fig:vertex_dep}.
We show how and if the transient currents (induced by the abrupt
bias switch at time $\tsw = 500\,\hbar/|\tB|$)
depend on chosen drain vertex site $n$.
In Fig.~\ref{fig:vertex_dep} we plot transient local currents obtained from 
our simulations for three subsequent rings sizes and several drain-lead
attachments.
These rings represent various different kinds of electronic transport properties
which can be found in rings.
The  local currents shown have been calculated at lattice site 120.
The finite basis set adds rapid oscillations on the computed local
currents\footnote{%
The period of these artificial oscillations
is always $\tau$, the quantity defined by eq.~(\ref{eq:tau_n}),
here equal to $\pi\,\hbar/|\tB|$ (same for both the bands).
The oscillations are larger at sites closer to the ring.
All this is explained by the localised nature of the stroboscopic basis states (the wavepackets)
and the finite number of them.
The train of the wavepackets (see Fig.~\ref{fig:packets})
travels in real space in such a way that during a time interval
$[t, t+\tau]$ each packet places itself exactly to the position which was occupied by its neighbour
at time $t$.
Since we use the finite cutoff on the basis set,
the quality of the description at given point of space fluctuates in time with the period of $\tau$.
Instead of using a huge value of $\mmax$ (which would be prohibitively expensive)
we smooth out the oscillations by averaging them
over the time interval of about $\tau$.
Typical timescales of the processes studied in the present work are in most cases
much larger than $\tau$ and the relevant effects are not affected by
the oscillations or by their artificial smoothening.}.
We smooth out the oscillations by averaging them over the time interval of about $\tau$.
This averaging does not modify essential features of the plots including 
the characteristic times and rates.
Example of the raw unaveraged data are shown in Fig.~\ref{fig:slope_vs_noslope}.
The ``no slope'' black curves in that figure
are obtained in the same system as plots $n=7$ and $n=8$ in
Fig.~\ref{fig:vertex_dep}, apart from the time averaging.

The onset of the electronic responses appears at times around
$550\,\hbar/|\tB|$, which is delayed about 50 time units after
$\tsw$, caused by the distance from the ring to the observation
site $l=120\ $ 
[$t_l \sim (l-N) a /v_\mathrm{F} \approx 50 \, \hbar/|\tB|\,$].
We see, that
similarly as in the stationary regime, the time-dependent 
simulations also 
exhibit the two very different regimes - the conducting and 
insulating one.
However there is the peak of transient current
even in the insulating 
configurations after the bias is switched on.
This feature is shown in the lower panel of Fig.~\ref{fig:CondInsu_SurfPlots}
and will be discussed also below.

$\boldsymbol{N=16}$ \textbf{rings.}
The left panel of Fig.~\ref{fig:vertex_dep} shows results for rings of
size $N=16$, but differing by the position of the drain vertex sites which run through
the sequence $n = 5, \dots, 8$.
The odd-numbered cases ($n = 5$ and $n = 7$) correspond to conducting configurations.
The well conducting state
of the ring is given by constructive interference (CI) of the electron
amplitudes in the two ring branches.
CI are most significant  in the case of symmetrically
attached ring $N=16$, $n=9$,
the configuration with equally long branches (not shown in the figure).
As we can see from the figure, and in agreement with former theoretical
analysis~\cite{Stefanucci09,SGML11}, significant CI is possible for
several vertex configurations.
The two local currents  ($n = 5$ and $n = 7$) reach similar long-time limits and the time
needed to build up the current is the same for each drain vertex site:
$\Delta t \approx 10\,\hbar/|\tB|$.
The dashed lines in Fig.~\ref{fig:vertex_dep} represent stationary
currents as obtained from the exact-eigenstate approach.
At intermediate times, around 600 time units,
the dynamically computed currents are in 
good agreement with the stationary approach.
The agreement would become slightly worse at larger times when finite basis set
errors take effect.
See also Fig.~\ref{fig:stat_curr} for comparisons of stationary currents
also at other biases.
On the other hand there are the even-numbered cases which show almost
zero currents after a short transient effect.
The peaks of the transient currents (the plots with $n=6$ and
with $n=8$) are very similar each other.
The blocking status is again due to quantum interference, now the destructive one.
Results for other values of $n$, not shown in plots, also confirm that
the transient characteristics are practically
the same for all drain vertex attachments within the particular group
(conductive or insulating).

$\boldsymbol{N=17}$ \textbf{rings.}
Central panel of Fig.~\ref{fig:vertex_dep} shows results for rings with
size $N=17$.
In this case there is no such a distinct separation into conductive
and insulating configurations,
the $N=17$ rings are all conductive.
Neither constructive nor destructive inteferences are now perfect.
However, similarly as in the insulating configurations
and contrary to the conducting configurations of
even-sized rings,
we now observe peaks in transient currents.
In addition there is a substantially longer relaxation, now lasting for
about 100 time units.
The peak current depends on the drain vertex site in an oscillatory
manner.
(More DC-conductive rings in this class exhibit lower
peak currents.)
$N=17$ rings have shown to be more difficult from computational point of view.
The effect of the incomplete basis set
would show up at long-time stationary currents which would be,
by estimate, 10-25\% lower compared to exact results from eigenstates.

$\boldsymbol{N=18}$ \textbf{rings.}
In this case (right panel of Fig.~\ref{fig:vertex_dep})
we again obtain two distinctly different behaviours.
The difference from the $N=16$ ring is that now the odd-numbered drain
vertices correspond to insulating configurations.
The long-time limit of the time-dependent treatment would relate 
to exact stationary results similarly as for $N=17$ rings.
In addition, the ring with $18$ atoms and with the right vertex at site $n=5$
exhibits pronounced circulating currents as will be discussed in separate 
subsection~\ref{ssec:circul}.

The results show, that for even-numbered rings, the transient effect
is very short, lasting for about $10$ time units.
Surprisingly, the odd-numbered ring $N=17$ exhibits longer relaxation,
lasting for about $100$ time units, although the initial development
upon the bias switch is equally rapid as for $N=16$ and $N=18$ rings.
The long-time asymptotic values
of currents from our simulations exhibit either conducting or insulating
character which is in many cases in a good agreement
with the analytical stationary results.
%
%
\subsubsection{\label{sssec:slope}
Role of potential-energy profile within the ring}
%
The time evolution of the system in our model is considered in
the independent-electron approximation.
%
\begin{figure}[t]
\centerline{\includegraphics[width=86mm]{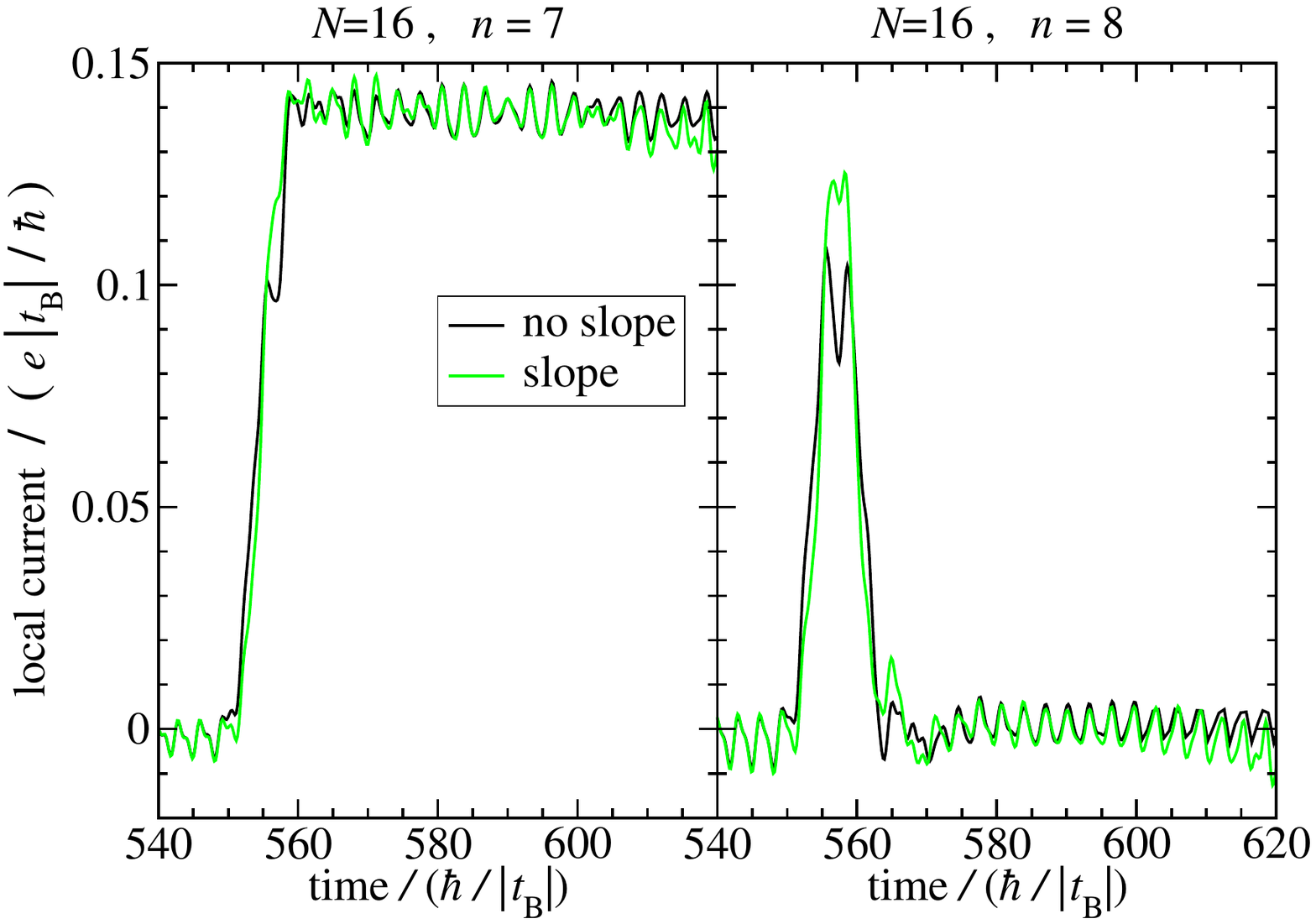}}
\caption{Transient local currents through the system with
$N=16$ ring computed in the drain lead (again at site $l = 120$)
for one conducting ($n = 7$) and one insulating ($n = 8$)
configuration and for two models of on-site energy
setting within the ring:
the model with spatially constant profile (``no slope'', black line)
and the model with linearly varying energies (``slope'', green or grey line).
The abruptly turned-on bias has the magnitude $0.5\,|\tB|/e$.
Basis set size for these plots are the same
as on Fig.~\ref{fig:vertex_dep}.
In contrast to other figures, the shown currents were not averaged over short time scales.
}
\label{fig:slope_vs_noslope}
\end{figure}
%
The on-site energies within
the ring are all set to the half of the instant bias value.
The on-site energy profile within the ring is spatially constant.
One may expect that the results would be different if we had considered
spatial variation of the on-site energies within
the ring.
The energies might, for example, be computed on-the fly as dependent on 
the electron charge density which would be a simple dynamical mean-field model.
Another model for ring on-site energies may be their linear variation (slope)
from one terminal of the ring to the opposite terminal such that the
on-site energies of the vertex atoms reach those in the corresponding leads.
We checked both these models and found that they do not have any significant
impact on the electron dynamics in the tested cases.
This is illustrated in Fig.~\ref{fig:slope_vs_noslope} for
a ring composed of 16 atoms and
the drain vertex being either at site
$n=7$ (conducting configuration)
or at site $n=8$ (insulating configuration).
As we can see the differences between the spatially uniform and the spatially
varying model are essentially negligible for both conducting and insulating
configurations.
This finding may not be universally valid.
However,
to keep the models in the present work simple, we stick at the spatially uniform
profile of on-site energies within the studied rings, with obvious exceptions
of an applied gate potential (see Sec.~\ref{ssec:Vgate}).
%
%
\subsection{\label{ssec:circul}Circulating currents}
%
%
It has been discussed in several studies (e.g.~\cite{Stefanucci09})
that circulating currents (CCs) may arise in certain ring configurations.
These analysis were done for stationary currents.
In this subsection we study circulating currents in transient
regime upon the abrupt switch of the bias.
By definition, a CC at given instant of time in a ring structure (like that on
Fig.~\ref{fig:ring}) exists when the currents
in the two ring branches have opposite directions in the sense that, for example,
the electrons in the shorter branch flow from the left terminal to the right one while
the electrons in the longer branch flow from the right terminal to the left one.
The currents in the ring branches are calculated using formulae~(\ref{eq:current_total_strobo})
and (\ref{eq:current_1ele_strobo}).
These formulae represent strictly local currents:
$I_{l_0}(t)$ is in fact the current between sites $l_0$ and $l_0+1$,
i.e. a current through the bond.
Our calculation of the current in given ring branch in addition involves
averaging over the interatomic bonds of given branch.

Inspection of the currents calculated from exact transmittances
of Ref.~\cite{SGML11}
shows that stationary CCs are possible for many ring sizes.
Actual occurrence of the CCs depends also on chosen drain site
and on applied bias.
Of the studied rings we found most pronounced CC in 
$N=18$ ring with its drain vertex at site $n=3$ and $n=5$.
%
\begin{figure}[!t]
\centerline{\includegraphics[width=86mm]{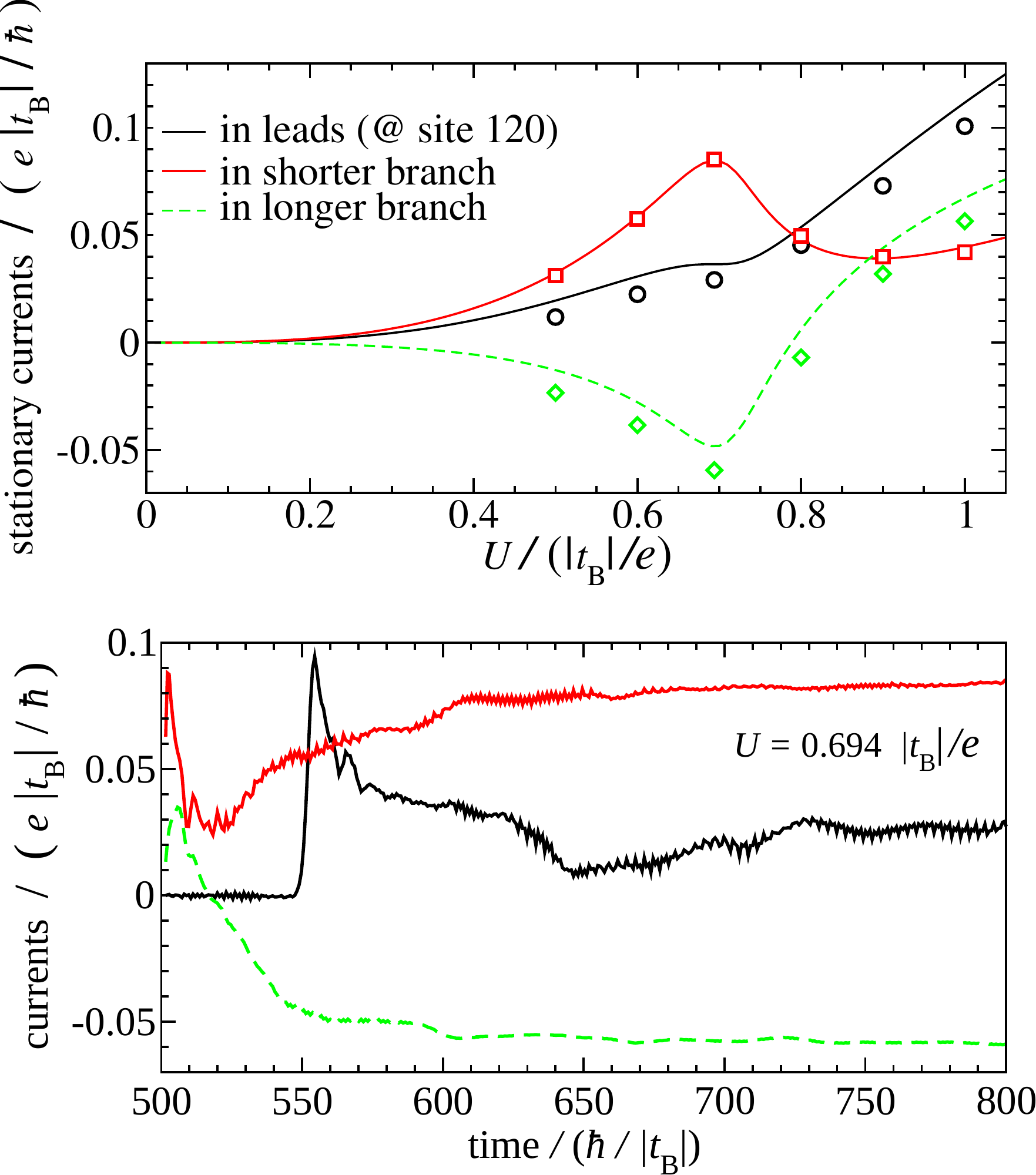}}
\caption{Electron currents in the system lead and in the two ring branches,
computed for $N=18$, $n=5$ ring.
\textbf{Upper panel:} Line plots show stationary currents computed
from exact transmittances~\cite{SGML11}.
Point symbols show quasi-stationary values obtained in our simulations
in a long-time limit (times about 1000 units).
Basis set corresponding to $\mmax = 1200$ (larger than in other figures)
has been used.
\textbf{Lower panel:} Time-dependent electron currents
induced by abrupt bias switch at time $t_\mathrm{sw} = 500\,\hbar/|\tB|$,
computed for bias $0.694\,|\tB|/e$.
This value of bias corresponds to the largest circulating
current in the stationary regime.
Colour and line-style coding is the same as on the upper panel.
Currents in the branches are spatial averages over sites within particular
branch (excluding the vertex sites).
Currents in the right lead (computed at site 120) have been smoothened
similarly as described in caption to Fig.~\ref{fig:vertex_dep}.
$\mmax = 1200$ has been employed also here.
}
\label{fig:circul_curr}
\end{figure}
%
The results for the $n=5$ structure are shown in Fig.~\ref{fig:circul_curr}.

Upper panel in Fig.~\ref{fig:circul_curr} shows the exact stationary electron 
currents, calculated with the approach described in~\cite{SGML11} (lines),
alongside with the quasi-stationary values from our simulations 
computed with the basis-set parameter $\mmax=1200$ (isolated symbols)~\cite{explain_SGML}.
Red (dark grey solid line) and green (light grey dashed line) plots display currents in the two ring branches and are relevant 
in verifying if CC is present.
The sign convention used on the figure is such that currents in both branches have positive
signs for the flow of the particles from left to right,
i.e. state without CCs.
The negative value of one of the currents indicates the presence of the CC in the ring.
From the upper panel of Fig.~\ref{fig:circul_curr} we see that
such a negative electron current flows in stationary regime in the longer ring branch for a wide range of
voltages (0 to $0.78~|\tB|/e$).
Most pronounced CC is found at a bias of $U = 0.694\,|\tB|/e$.

Lower panel of Fig.~\ref{fig:circul_curr} extends the results
at the bias $0.694\,|\tB|/e$ into the non-stationary regime.
The non-stationarity is due to the abrupt bias switch at time
$t_\mathrm{sw} = 500\,\hbar/|\tB|$.
The colour and sign convention is the same as for the upper panel.
Negative currents show up in the longer ring branch.
Black plot on the lower panel of Fig.~\ref{fig:circul_curr}
represents current computed at site 120 (i.e. the usual current in the right
lead at a position 115 sites beyond the right vertex atom)\footnote{%
The smoothening of the artificial oscillations (see previous footnote)
is not always perfect as can be seen
from the bottom panel of Fig.~\ref{fig:circul_curr}, especially the black and red plots.
This is a purely technical issue caused by a too large interval between recorded data from the simulation
and by the width of the window used to compute the running averages.
It is interesting that the green plot in the graph does not show the oscillations.
In fact they are present but are very small in magnitude.
The impact of the finite basis set on the presence of the oscillations is often quantitatively
different in different parts (leads and ring branches) of the whole system.}.

Finally, the dip in the right-lead current (black plot on
the lower panel of Fig.~\ref{fig:circul_curr})
at times ranging from about 630 to $730\,\hbar/|\tB|$
is caused by the numerical cutoff on the stroboscopic basis set size.
This unphysical dip tends to be more pronounced in certain ring configuration,
for certain quantities and for high biases.
Ring configuration $(N,n) = (18,5)$ is significantly affected by the
finite basis set error at certain time interval.
This is also the reason why we used parameter $\mmax=1200$
to obtain results shown in Fig.~\ref{fig:circul_curr}
while for most of other figures we use $\mmax=352$.
The increased cutoff improves accuracy of time-dependences especially
at intermediate times.
%
%
\subsection{\label{ssec:Vgate}Time-dependent currents controlled by gate field}
%
%
Quantum interference effects between electrons in the two ring arms
can be controlled by an applied electric field.
The field value can be time-dependent in general.
In our approach it is modelled by a spatially uniform lift of on-site energies
in one of the branches of the ring (not including the two terminal atoms);
the affected atoms have indices $n+1$, $n+2$, \dots, $N$ in the indexing scheme of Fig.~\ref{fig:ring}.
\subsubsection{Abrupt gate turn-on}
\begin{figure}[t]
\centerline{\includegraphics[width=86mm]{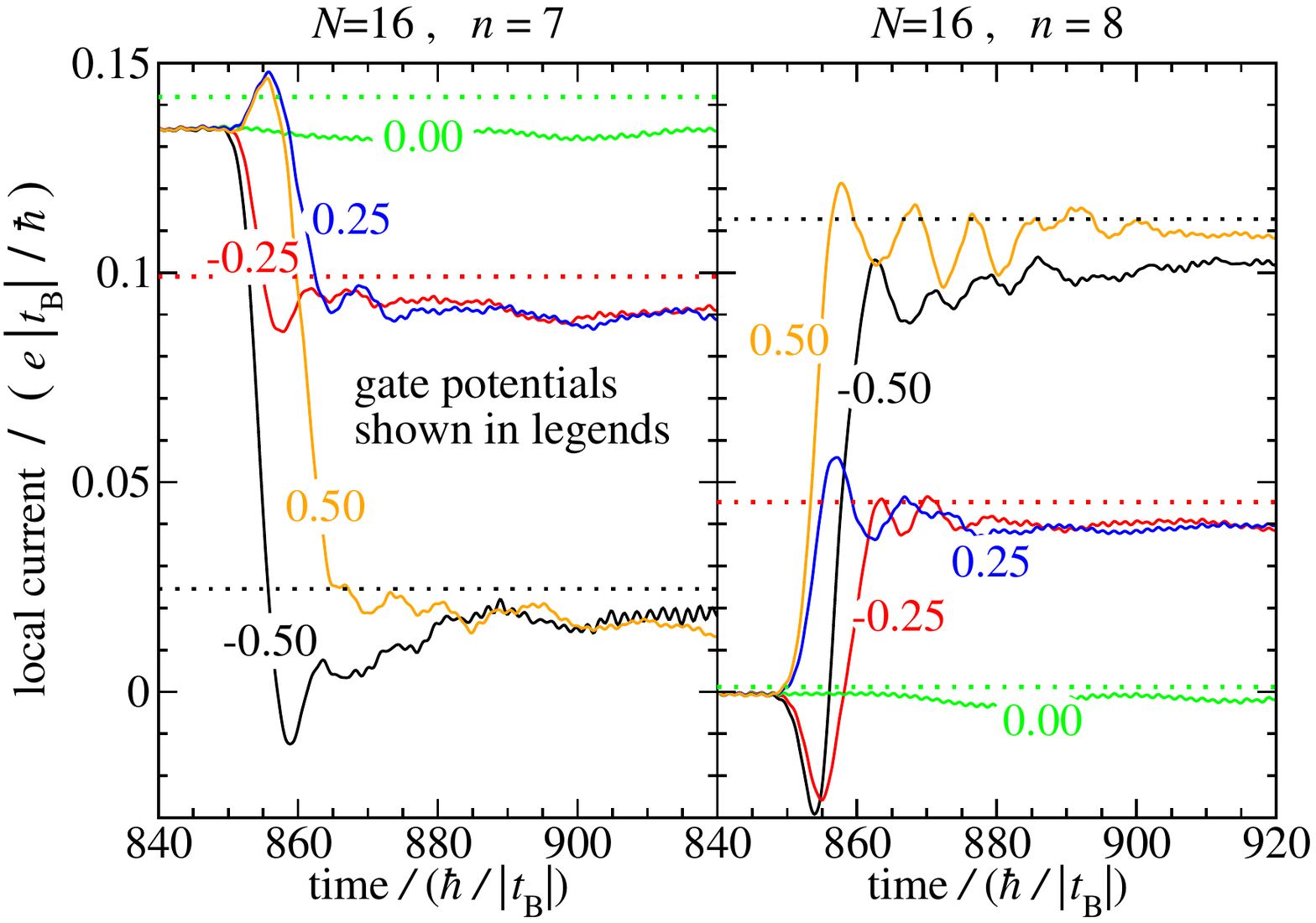}}
\caption{Local electron currents computed at site 120 when gate field
is abruptly turned on at time $t_\mathrm{g} = 800\,\hbar/|\tB|$.
The applied bias is constants in these plots and equal to $U = 0.5\,|\tB|/e$.
Values of the gate potential are shown in legends and use the same units as
the bias.
The ring has $N = 16$ atoms.
The plots in the left panel use drain vertex attached at site
$n = 7$ (conducting configuration)
while the plots in the right panel use $n = 8$ (insulating 
configuration).
The dotted lines represent analytically computed stationary currents as obtained
by the exact-eigenstate method described in Sec.~\ref{sec:eigenstates}.
The exact stationary currents are even functions of the gate potential.
}
\label{fig:Vgate_abrupt}
\end{figure}

In our first studied application of the gate potential, 
we control the electron transport using an instantaneously switched gate field
which is applied to the two $N=16$ rings differing by their drain terminal indices $n$.
The field is turned on at time $t_\mathrm{g} = 800\,\hbar/|\tB|$, after the 
system has been evolving under the constant bias of $U = 0.5\,|\tB|/e$ so that 
it is essentially in a stationary-current regime before the time when the gate field
is lifted.
The simulations give results which are depicted
in Fig.~\ref{fig:Vgate_abrupt}.
%
%
The local current is evaluated at the lattice site 120.
We observe that the gate field induces transient effects lasting for 
about $10-40$ time units.
The transient effects depend rather strongly on particular bias values.
The application of the higher gate field is interesting;
it would allow the ring to work as a field driven switching device,
i.e. a nanoscale-sized transistor,
changing the conducting ring into non-conducting or vice versa.
The switching is a consequence of effective change in Fermi wavelength
of electrons in the controlled (longer) branch and hence changing the interference
pattern which shows up in the resulting current.
The dynamically computed currents at longer times agree well with the exact 
values from eigenstates although differences arise especially at higher 
gate potentials.
As in other calculations, the source of the differences is
the finite cutoff on the number of stroboscopic basis packets used.
\subsubsection{Harmonic gate potential}
\begin{figure}[t]
\centerline{\includegraphics[width=86mm]{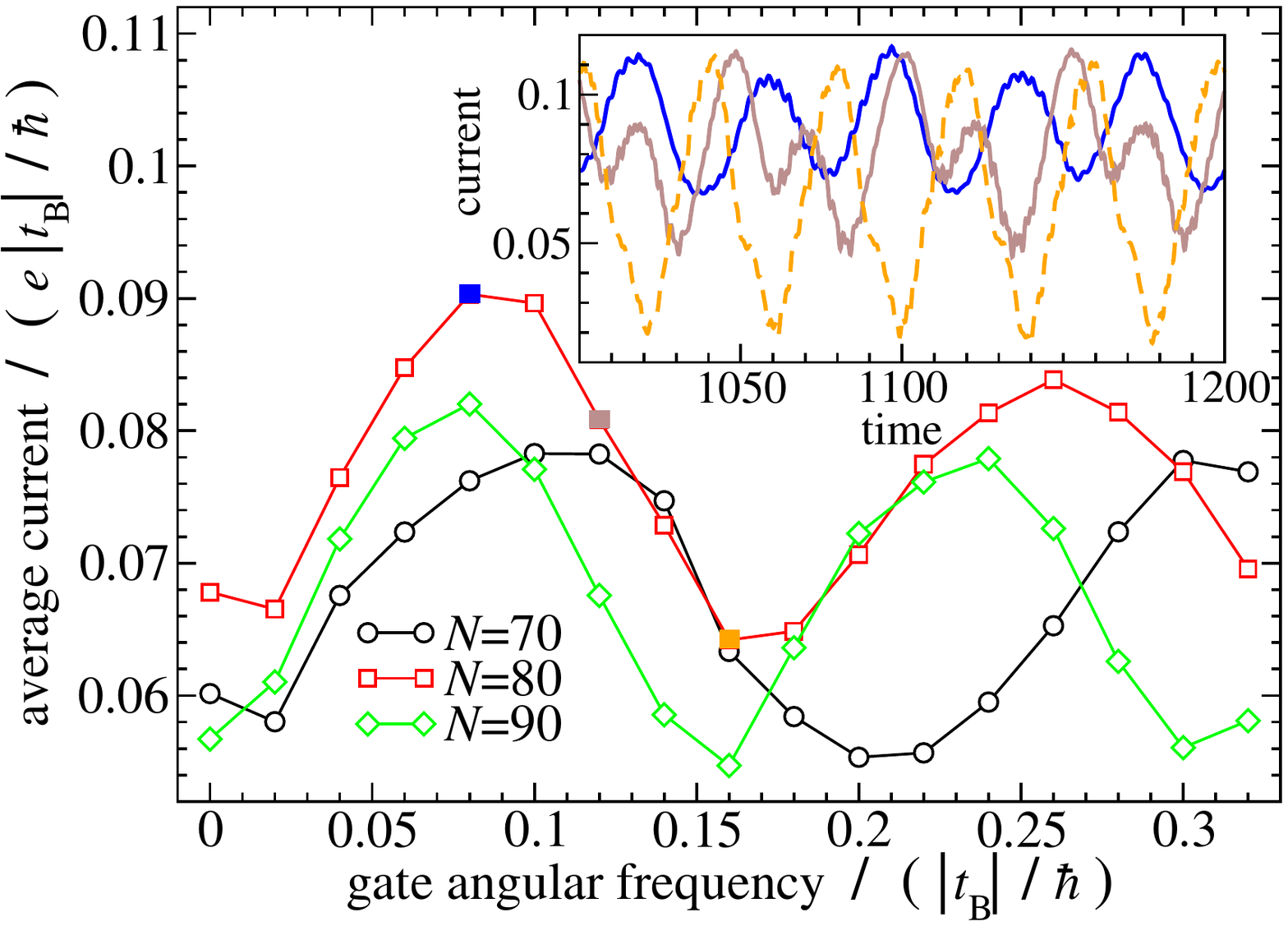}}
\caption{\textbf{Main graph:} Average lead currents as functions of the sinusoidal gate
potential for three different ring structures:
$N=70$, $n=61$ (black line with circles), $N=80$, $n=71$ (red line with squares),
$N=90$, $n=81$ (green line with diamonds).
The bias is always $U = 0.5\,|\tB|/e$ (time-independent).
The gate field acts on the atoms $n+1$, \dots, $N$.
The amplitude of the gate potential is always $V_\mathrm{g0} = 0.25\,|\tB|/e$.
\textbf{Inset:} Detailed time-dependences shown for three particular points from the red plot
of the main graph: the case of $N=80$ with $\omega_\mathrm{g} = 0.08$ (dark blue plot, solid line),
$0.12$ (brown plot, solid line), and $0.16\,|\tB|/\hbar$ (orange plot, dashed line).
Technical and basis set parameters used for these results are the same as for most
of the other plots; in particular, $\mmax = 352$.
}
\label{fig:Vgate_oscil}
\end{figure}

Another option to control the current flow is to employ
a harmonically oscillating gate potential.
For this purpose we decided to use larger rings with sizes $N$ ranging from 70 to 90.
Such rings provide longer characteristic times of temporal changes.
We study the following three $(N,n)$ configurations: $(70,61)$, $(80,71)$, and $(90,81)$.
All the three rings thus have one longer and one shorter branch.
The gate field is again applied to atoms with indices in the range $n+1$, \dots, $N$,
i.e. the atoms of the shorter branch now.
The constant bias is again $U = 0.5\,|\tB|/e$.
The magnitude of the sinusoidal gate potential is always $V_\mathrm{g0} = 0.25\,|\tB|/e$.
The potential is harmonic with various angular frequencies as shown on Fig.~\ref{fig:Vgate_oscil}.
%
%
The potential starts to act at time $t_\mathrm{g} = 800\,\hbar/|\tB|$.
The inset of Fig.~\ref{fig:Vgate_oscil} shows typical time dependencies of the current
in leads as obtained for the case $(N,n) = (80,71)$.
The three plots of the inset have been recorded for three different gate angular frequencies
$\omega_\mathrm{g}$: $0.08$ (dark blue, solid line), $0.12$ (brown, solid line)
and $0.16$ of $|\tB|/\hbar$ (orange, dashed line).
The corresponding periods $T_\mathrm{g} = 2\pi/\omega_\mathrm{g}$ are approximately $78.5$,
$52.4$, and $39.3$ of $\hbar/|\tB|$.
The current response is generally quasi-periodic but anharmonic.
(The initial transient effect after the gate is turned on is not discussed here.)
The temporal dependence of the current is given by an interplay of the harmonic gate potential 
and internal ring effects given also by its size.
At low gate frequencies $\omega_\mathrm{g}$ (for example $0.08\,|\tB|/\hbar$ or less,
see the blue plot, dark solid line, in the inset),
the current oscillations typically (not always) exhibit periodic quasi-harmonic
pattern at the twice of the gate frequency.
The doubled frequency arises from the symmetric dependence of the system transmittance on the
gate potential.
($V_\mathrm{g}$ and $-V_\mathrm{g}$ have the same effect on stationary currents as could be
seen from the analytic formulae of Ref.~\cite{SGML11} or calculated by formalism of Sec.~\ref{sec:eigenstates}.)
Rapid driving is not followed by the current in this sense as can be seen from the inset.
For example, the gate field with $\omega_\mathrm{g} = 0.16\,|\tB|/\hbar$
(the orange plot with dashed line on the figure)
or at higher gate frequencies results in current oscillations at the same frequency.
Intermediate driving frequencies (e.g. $0.12\,|\tB|/\hbar$, the brown plot, solid line, on the figure)
yield periodic but anharmonic time dependences.

It is interesting to compute time averages from the oscillating lead currents.
The three plots in the main graph of Fig.~\ref{fig:Vgate_oscil} show that
the average current for given ring structure depends on the gate frequency
(while the bias and the gate amplitude are kept unchanged).
The maxima and minima of the average currents vary with varying ring structure:
larger rings have the extrema shifted towards lower frequencies.
Inspection of the plots show that these variations fulfil the law
$\delta\omega/\omega = -\delta N/N$ which is expected from elementary considerations
about resonant frequencies.
In this way we have an evidence that the oscillatory pattern of the
average current plotted in Fig.~\ref{fig:Vgate_oscil} arises from the 
internal ring resonances.
Given the nanometer-scale size of such ring structures, the characteristic
resonant frequencies lie in the optical domain and we do not investigate
higher gate frequencies that those shown in the graph.

Electric current dynamics could be investigated also for a fixed ring size $N$
and varying drain terminal $n$.
However, our inspection has shown that the dynamics is more interesting 
(i.e. the average currents exhibit more pronounced oscillations as functions
of $\omega_\mathrm{g}$)
when the source and drain terminal are relatively close to each other.
%
%
\subsection{\label{ssec:coupling}Reduced coupling to the leads}
%
%
Through previous sections it was assumed that all nearest neighbour 
couplings were identical along the whole composed system, i.e. including the 
small system (the ring).
The couplings between the leads and the ring were then quantified
by the hopping parameter $\tB < 0$ (same as in the lead and in the ring).
%
%
\begin{figure}[t]
\centerline{\includegraphics[width=86mm]{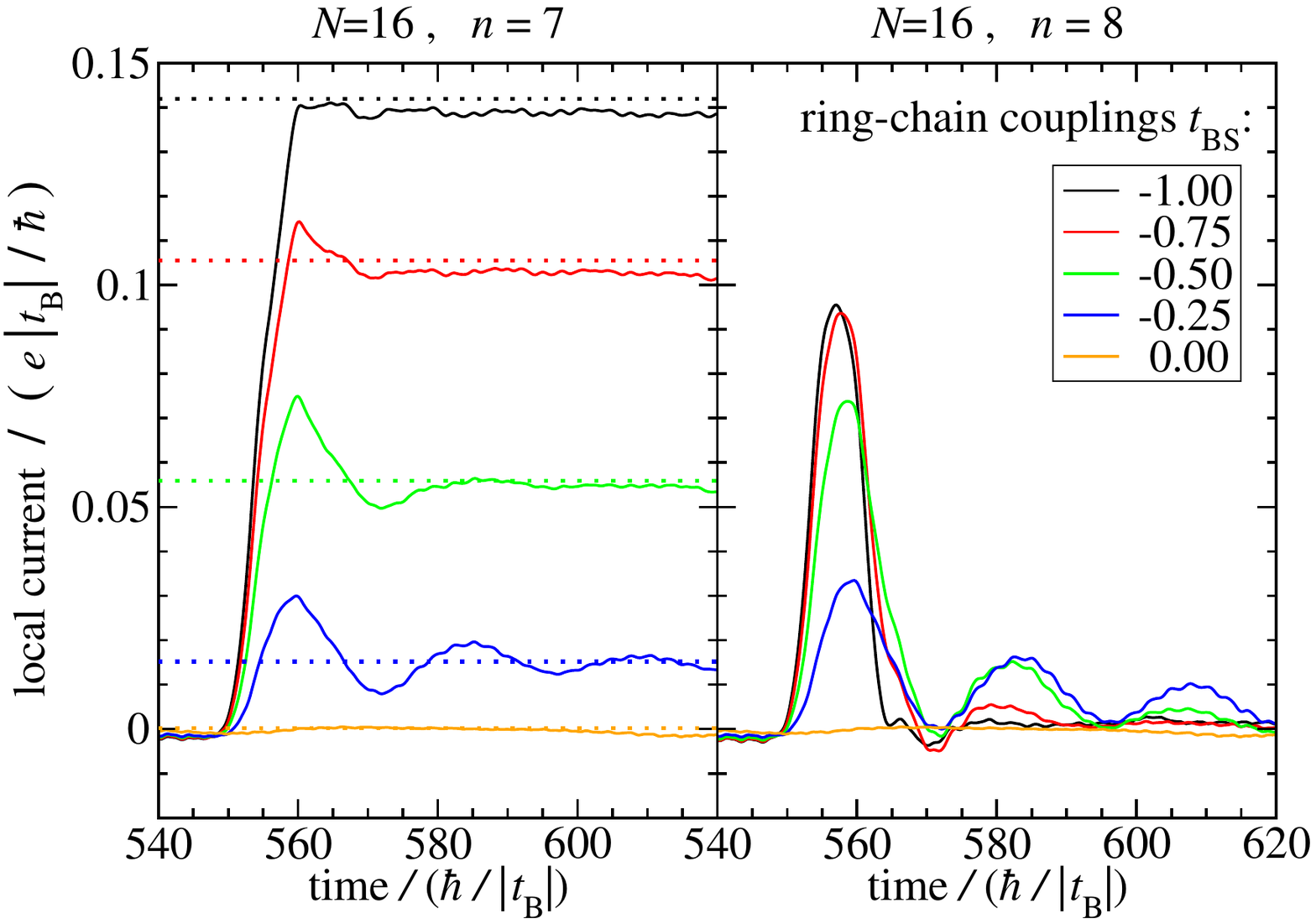}}
\caption{Local electron currents evaluated at site 120
for various lead-system coupling strengths $\tBS$.
Legends indicate values of $\tBS$ relative to the magnitude of the
lead's hopping parameter $\tB$.
As in previously shown results, bias $U = 0.5\,|\tB|/e$
is abruptly turned on at time $t_\mathrm{sw} = 500\,\hbar/|\tB|$.
Number of ring atoms is $N=16$ in both panels and drain is attached to sites
$n=7$ (left panel) and $n=8$ (right panel).
The dotted lines in the left panel (conducting configuration) show stationary
currents computed from the exact-eigenstate approach.
}
\label{fig:coupl_dep_at_bias_dot5}
\end{figure}
%
In this section we study a more general situation when the couplings within 
the leads again remain equal to $\tB$ and the same parameter is used also
to bound the atoms included in the ring.
The couplings between the leads and the ring will however be characterised by
a new parameter $\tBS < 0$ (lead-system coupling).

We will report on regimes with $|\tBS| \le |\tB|$.
In the limiting case of $|\tBS| \to 0$ the leads and the system (the ring) become decoupled
and the internal energy levels of the ring attain discrete values.
In intermediate cases of $0 < |\tBS| < |\tB|$ we expect broadening
of the ring levels.
The internal ring electronic structure will have an impact on the transport
characteristics.

An illustration of our findings is shown on
Fig.~\ref{fig:coupl_dep_at_bias_dot5}.
The time-dependent currents have been computed for the rings of $N=16$ atoms,
left panel corresponding
to drain vertex site $n=7$ (conducting configurations) while
right panel shows plot for $n=8$ (insulating configurations).
In all cases the bias $U=0.5\,|\tB|/e$ was abruptly switched on at time
$t_\mathrm{sw}=500\,\hbar/|\tB|$.
The magnitude of the transient effect as well as the currents (in the conductive
system) behave as expected regarding to the lead-system coupling strength.
The insulating ring retains its property also at reduced couplings.
However, at intermediate coupling range $\tBS \in [-0.50$, $-0.25]$
there are decaying oscillations in the electron current persisting for 
a significant period of time, thus extending the transient effect to 
a period of about $100\,\hbar/|\tB|$.
Inspection shows that angular frequency $\omega$ of these decaying
oscillations is related to
the applied bias by the relation $\hbar\omega = e U/2$, in agreement with 
previous analytical~\cite{bokes_PCCP} as well as numerical~\cite{BWang10} results.
%
%
\section{Conclusions}
%
%
Recently proposed stroboscopic wave packets~\cite{bokes_PRL,bokes_PCCP}
have been developed to be applicable to systems employing explicit
atomistic level of modelling.
Time-dependent transport of electrons in an open
system with localised perturbation was studied.
The localised perturbation had a ring structure described in
the tight-binding approximation.
Such a system is of interest due to quantum interference effects
that affect its transport properties.
It might serve as a field driven quantum interference
current switching device~\cite{Cardamone06} -- a nanoscale-sized transistor.
We have demonstrated the potential of our newly developed method based 
on the unitarily propagating stroboscopic wave packets to describe
open systems with fluctuating number of explicitly included electrons
while the whole system has an infinite number of electrons which can be
neglected in a good approximation.
In our method
full quantum coherence is preserved throughout the whole infinite system.
The method can be used for systems with one-dimensional semi-infinite leads.
It is capable of spatially resolved description of
transient effects like those caused by an abrupt bias switch
of a gate field application.
However, its most useful application would be found in systems which are 
exposed to long-lasting varying external fields or biases and 
for which it is important to preserve quantum coherence in the
description.
For short-time simulations one could use a less expensive
model with cyclic boundary conditions.
However, such an approach would become prohibitively expensive
for large simulation times as it would require to consider many
explicit atoms to describe the leads.
In contrast, our method does not have any such limit on simulation
time.
The weakness of the present implementation of the stroboscopic basis set description
is slow convergence
of relevant results with increasing number of basis functions.
The finite basis set demonstrates itself in a transient unphysical drop
of lead current in some of the studied systems.
Depending on the studied configuration, long-time stationary currents
may also sometime be significantly underestimated.
Other consequence of the finite stroboscopic basis set are small artificial rapid oscillations
in computed quantities which can however be smoothed out and usually does not prevent us 
from capturing relevant physical time-dependent effects.
In our ongoing work we will consider a generalisation of the basis
set in order to reduce the finite-basis set errors and to reduce the number
of basis states needed.
%
%
%
%
\section{Acknowledgements}
%
%
This work was supported in parts by
the Slovak Grant Agency for Science (VEGA) through grant No.~1/0632/10
and by
the Slovak Research and Development Agency under the contract No.~APVV-0108-11.
\appendix
%
%
\section{\label{app:ovlp} Derivation of the $\langle l|n,\alpha,m;t\rangle$ overlaps}
%
%
Here we derive formula~(\ref{eq:overlap}) for the overlaps between the TB atomic orbitals
$|l\rangle$ and the SWPs $|o;t\rangle \equiv |n,\alpha,m;t\rangle$.
In the SWPA~\cite{bokes_PRL,bokes_PCCP}, the eigenstates $|\mathcal{E},\alpha\rangle$
of the basis-generating Hamiltonian $\hat{H}^0$ [given by eq.~(\ref{eq:hamilton_TB})
in this work] use the energy ``normalisation'' condition which is
\begin{equation}
\langle \mathcal{E},\alpha|\mathcal{E}',\alpha'\rangle
=
\delta(\mathcal{E}-\mathcal{E}') \delta_{\alpha,\alpha'}
\ .
\label{eq:norm}
\end{equation}
See also sec.~\ref{sec:strobo} and eq.~(\ref{eq:basis_vec}) therein.
This condition leads to the form
\begin{equation}
|\mathcal{E},\alpha\rangle
=
\frac{1}{\sqrt{\left|\frac{\partial\mathcal{E}}{\partial k}\right|}} \, |k, \alpha\rangle
\end{equation}
where $|k, \alpha\rangle$ is the usual Bloch wave in one dimension with the k-number of magnitude $k$ and
the propagation direction labelled by $\alpha = \pm 1$.
Normalisation of the Bloch waves is assumed to be
\begin{equation}
\langle k,\alpha|k',\alpha'\rangle = \delta(k-k') \delta_{\alpha,\alpha'}
\ .
\end{equation}
(The k-numbers will be restricted to the interval
$k = [0, \, \pi/a]$ where $a$ is the lattice constant.)
In case of the TB model with the set of the orthonormal atomic orbitals
$|l\rangle$ (one orbital per atom) we obtain formula
\begin{equation}
|\mathcal{E},\alpha\rangle
=
\frac{1}{\sqrt{2 |\tB| \sin\mathcal{K}}} \, \sqrt{\frac{1}{2\pi}} \,
\sum_{-\infty}^{\infty} e^{i \alpha \mathcal{K} l} |l\rangle
\label{eq:Bloch_Enorm}
\end{equation}
with the summation running over all lattice sites.
$\mathcal{K} \equiv k a$ is the dimensionless wavenumber.
The eigenstates $|\mathcal{E},\alpha\rangle$ are used to construct the SWPs
according to eq.~(\ref{eq:basis_vec}).
Now we use that formula and write down the overlaps in the form
\begin{equation}
\langle l |n,\alpha,m;t\rangle
=
\frac{1}{\sqrt{\Delta\mathcal{E}_n}}
\int_{\mathcal{E}_{n-1}}^{\mathcal{E}_n}
\exp\left[-\frac{i}{\hbar} (m \tau_n + t) \mathcal{E}\right]
\langle l|\mathcal{E},\alpha\rangle \,
\dd\mathcal{E}
\label{eq:ovlp_aux1}
\end{equation}
where we have utilised the equation
\begin{equation}
\hat{H}^0 |\mathcal{E},\alpha\rangle = \mathcal{E} |\mathcal{E},\alpha\rangle
\ .
\end{equation}
Using the expression~(\ref{eq:Bloch_Enorm}) for the eigenstates and substituting it into
eq.~(\ref{eq:ovlp_aux1}) leads to an  integral expression with integration variable
$\mathcal{E}$.
With the aid of the TB dispersion relation~(\ref{eq:disprel}), the integral can be
transformed to the integration variable $\mathcal{K}$.
The final form of the expression for the overlap $\langle l |n,\alpha,m;t\rangle$
is then provided by formula~(\ref{eq:overlap}).
We evaluate these integrals numerically.

\begin{thebibliography}{99}
%
\bibitem{NEGF}
H.~Haug and A.-P.~Jauho,
\textit{Quantum Kinetics in Transport and Optics of Semiconductors}
(Springer-Verlag, Berlin, Germany, 1998)
%
\bibitem{Ness11}
H.~Ness and L.K.~Dash, Phys. Rev. B \textbf{84}, 235428 (2011)
%
\bibitem{Cini80}
M.~Cini, Phys.~Rev.~B \textbf{22}, 5887 (1980)
%
\bibitem{Stefanucci04}
G.~Stefanucci and C.-O.~Almbladh,
Phys.~Rev.~B \textbf{69}, 195318 (2004)
%
\bibitem{TDDFT}
See for example
S.~Kurth, G.~Stefanucci, C.-O.~Almbladh, A.~Rubio, and E.K.U.~Gross,
Phys.~Rev.~B \textbf{72}, 035308 (2005)
and the references therein
%
\bibitem{MBPT}
P.~My\"{o}h\"{a}nen, A.~Stan, G.~Stefanucci, and R.~van~Leeuwen,
Phys. Rev. B \textbf{80}, 115107 (2009)
%
\bibitem{BWang10}
B.~Wang, Y.~Xing, L.~Zhang and J.~Wang,
Phys.~Rev.~B \textbf{81}, 121103 (2010)
%
\bibitem{YWang11}
Y.~Wang, C.Y.~Yam, G.H.~Chen, Th.~Frauenheim, and T.A.~Niehaus,
Chem.~Phys. \textbf{391}, 69 (2011)
%
\bibitem{Chen07}
X.~Zheng, F.~Wang, C.Y.~Yam, Y.~Mo, and G.~Chen,
Phys.~Rev.~B \textbf{75}, 195127 (2007)
%
\bibitem{Chen12}
H.~Xie, F.~Jiang, H.~Tian, X.~Zheng, Y.~Kwok, S.~Chen, C.~Yam, and G.~Chen,
J.~Chem.~Phys. \textbf{137}, 044113 (2012)
%
\bibitem{Torfason12}
K.~Torfason, A.~Manolescu, V.~Molodoveanu, and V.~Gudmundsson,
J.~Phys.: Conf.~Series \textbf{338}, 012017 (2012)
%
\bibitem{Hyldgaard12}
P.~Hyldgaard,
J.~Phys.: Condens. Matter \textbf{24}, 424219 (2012)
%
\bibitem{Horsfield04}
A.P.~Horsfield, D.R.~Bowler, and A.J.~Fisher,
J.~Phys.: Condens. Matter \textbf{16}, L65 (2004)
%
\bibitem{McEniry07}
E.J.~McEniry, D.R.~Bowler, D.~Dundas, A.P.~Horsfield, C.G.~S\'{a}nchez, and T.N.~Todorov,
J.~Phys.: Condens. Matter \textbf{19}, 196201 (2007)
%
\bibitem{McEniry10}
E.J.~McEniry, Y.~Wang, D.~Dundas, T.N.~Todorov, L.~Stella, R.P.~Miranda, A.J.~Fisher, A.P.~Horsfield, C.P.~Race, D.R.~Mason, W.M.C.~Foulkes, and A.P.~Sutton,
Eur.~Phys.~J. B \textbf{77}, 305 (2010)
%
\bibitem{bokes_PRL}
P.~Bokes, F.~Corsetti, and R.W.~Godby,
Phys.~Rev.~Lett. \textbf{101}, 046402 (2008)
%
\bibitem{bokes_PCCP}
P.~Bokes,
Phys.~Chem.~Chem.~Phys.~\textbf{11}, 4579 (2009)
%
\bibitem{Yi03}
J.~Yi, G.~Cuniberti, and M.~Porto,
Eur.~Phys.~J. B \textbf{33}, 221 (2003)
%
\bibitem{Goyer07}
F.~Goyer, M.~Ernzerhof, and M.~Zhuang,
J.~Chem.~Phys. \textbf{126}, 144104 (2007)
%
\bibitem{Pickup08}
B.T.~Pickup and P.W.~Fowler,
Chem.~Phys.~Lett. \textbf{459}, 198 (2008)
%
\bibitem{Stefanucci09}
G.~Stefanucci, E.~Perfetto, S.~Bellucci, and M.~Cini,
Phys.~Rev.~B \textbf{79}, 073406 (2009)
%
\bibitem{Xia92}
J.-B.~Xia,
Phys.~Rev.~B \textbf{45}, 3593 (1992)
%
\bibitem{SGML11}
R.E.~Sparks, V.M.~Garc\'{\i}a-Su\'{a}rez, D.Zs.~Manrique, and C.J.~Lambert,
Phys. Rev. B \textbf{83}, 075437 (2011)
%
\bibitem{saha10}
K.K.~Saha, B.K.~Nikoli\'{c}, V.~Meunier, W.~Lu, and J.~Bernholc,
Phys.~Rev.~Lett. \textbf{105}, 236803 (2010)
%
\bibitem{saha09}
K.K.~Saha, W.~Lu, J.~Bernholc, and V.~Meunier,
J.~Chem.~Phys. \textbf{131}, 164105 (2009)
%
\bibitem{Kiguchi08}
M.~Kiguchi, O.~Tal, S.~Wohlthat, F.~Pauly, M.~Krieger, D.~Djukic, J.C.~Cuevas, and J.M.~van~Ruitenbeek,
Phys.~Rev.~Lett. \textbf{101}, 046801 (2008)
%
\bibitem{Kiguchi12}
M.~Kiguchi, S.~Nakashima, T.~Tada, S.~Watanabe, S.~Tsuda, Y.~Tsuji, and J.~Terao,
Small \textbf{8}, 726 (2012)
%
\bibitem{Bai12}
M.~Bai, J.~Liang, L.~Xie, S.~Sanvito, B.~Mao, and S.~Hou,
J.~Chem.~Phys. \textbf{136}, 104701 (2012)
%
\bibitem{Hong12}
W.~Hong, H.~Li, S.-X.~Liu, Y.~Fu, J.~Li, V.~Kaliginedi, S.~Decurtins, and T.~Wandlowski,
J.~Am.~Chem.~Soc. \textbf{134}, 19425 (2012)
%
\bibitem{nanoribbon_strips}
X.~Li, X.~Wang, L.~Zhang, S.~Lee, and H.~Dai,
Science \textbf{319}, 1229 (2008)
%
\bibitem{nanoribbon_FET}
X.~Wang, Y.~Ouyang, X.~Li, H.~Wang, J.~Guo, and H.~Dai,
Phys.~Rev.~Lett. \textbf{100}, 206803 (2008)
%
\bibitem{Reed1997}
M.A.~Reed, C.~Zhou, C.J.~Muller, T.P.~Burgin, and J.M.~Tour,
Science \textbf{278}, 252 (1997)
%
\bibitem{Huant06}
B.~Hackens, F.~Martins, T.~Ouisse, H.~Sellier, S.~Bollaert, X.~Wallart, A.~Cappy, J.~Chevrier,
V.~Bayot, and S.~Huant,
Nature Physics \textbf{2}, 826 (2006)
%
\bibitem{Huant07}
F.~Martins, B.~Hackens, M.G.~Pala, T.~Ouisse, H.~Sellier, X.~Wallart, S.~Bollaert, A.~Cappy, J.~Chevrier,
V.~Bayot, and S.~Huant,
Phys.~Rev.~Lett. \textbf{99}, 136807 (2007)
%
\bibitem{Wieck10}
W.~Lei, C.~Notthoff, A.~Lorke, D.~Reuter, and A.D.~Wieck,
Appl.~Phys.~Lett. \textbf{96}, 033111 (2010)
%
\bibitem{Cardamone06}
D.M.~Cardamone, C.A.~Stafford, and S.~Mazumdar,
Nano~Lett. \textbf{6}, 2422 (2006)
%
%
\bibitem{bands}
Stroboscopic wavepacket method can in general work with arbitrary bulk
(or even non-periodic) Hamiltonian.
The bands need not be two and they need not be equally wide.
Our implementation is quite general with respect to the division of 
Hamiltonian energy spectrum into this kind of bands.
In all calculations in this work we however use two equally wide bands because
such setup is most practical.
%
%
\bibitem{explain_ElRemove}
In actual simulations we use a rather conservative criterion for electron removal.
Consequently, no electron is removed during simulated timescales.
%
\bibitem{e-e}
Studies of e-e interactions in connection with quantum interference effects 
have been published, see for example
P.~Stefa\'{n}ski, J.~Phys.: Condens.~Matter \textbf{22}, 505303 (2010).
%
\bibitem{numrec}
W.H.~Press, S.A.~Teukolsky, W.T.~Vetterling, and B.P.~Flannery,
\textit{Numerical recipes in C}, 2nd edn.
(Cambridge University Press, Cambridge 1994)
%
\bibitem{explain_KL}
For example, we have $\mathcal{E} = \eL + 2\tB \cos\KL$ in the left lead
and $\mathcal{E} = \eS + 2\tB \cos\KS$ in the ring.
%
%
\bibitem{I_in_ring}
Inspection of the currents through branches of unsymmetrical rings shows that
the situation is more complicated inside such rings.
See Fig.~\ref{fig:circul_curr} as an example.
%
%
\bibitem{explain_SGML}
Here we could equivalently use our exact-eigenstate method described in Sec.~\ref{sec:eigenstates}.
We instead used formulas of Ref.~\cite{SGML11} because of their convenience.
%
%
\end{thebibliography}
\end{document}